\documentstyle[12pt,epsfig]{article}

\newcommand{\lsim}{\raisebox{-0.13cm}{~\shortstack{$<$ \\[-0.07cm] $\sim$}}~}
\newcommand{\gsim}{\raisebox{-0.13cm}{~\shortstack{$>$ \\[-0.07cm] $\sim$}}~}

\newcommand{\ra}{\rightarrow}

\newcommand{\tb}{\tan \beta}
\newcommand{\s}{\smallskip}
\newcommand{\nn}{\noindent}
\newcommand{\non}{\nonumber}
\newcommand{\beq}{\begin{eqnarray}}
\newcommand{\eeq}{\end{eqnarray}}

\begin{document}

\baselineskip=17.pt

\begin{flushright}
CERN TH/2002--375\\
PM/03--02\\
UFIFT-HEP-03-3\\
TIFR/HEP/03-02
\end{flushright}

\vspace*{0.9cm}

\begin{center}

{\large\sc {\bf Detection of MSSM Higgs bosons from supersymmetric particle}}

\vspace*{0.2cm}

{\large\sc {\bf cascade  decays at the LHC}}

\vspace{0.5cm}

{\sc Aseshkrishna DATTA$^{1}$, Abdelhak DJOUADI$^{2,3}$,} 

\vspace{0.2cm}

{\sc  Monoranjan GUCHAIT$^{4}$} and {\sc  Filip MOORTGAT$^5$}

\vspace{0.5cm}

$^1$ Institute of Fundamental Theory, Department of Physics, \\ 
University of Florida, Gainesville, FL 32611,  USA. \s

$^2$ Theory  Division, CERN, CH--1211 Geneva 23, Switzerland.\s

$^3$ Laboratoire de Physique Math\'ematique et Th\'eorique, UMR5825--CNRS,\\
Universit\'e de Montpellier II, F--34095 Montpellier Cedex 5, France. \s

$^4$ Department of High Energy Physics, Tata Institute of Fundamental 
Research, \\  Mumbai-400005, India. \s

$^5$ Department of Physics, University of Antwerpen, \\ 
Universiteitsplein 1, B--2610 Antwerpen, Belgium.
\end{center} 

\vspace*{.5cm} 

\begin{abstract} 

\nn In the context of the Minimal Supersymmetric extension of the Standard
Model, we study the production of Higgs bosons at the Large Hadron Collider 
via cascade decays of scalar quarks and gluinos. We focus on the cascades
involving heavier charginos and neutralinos, which decay into the neutral
$h,A,H$ and charged $H^\pm$ bosons and lighter charginos and neutralinos, but
we will also  discuss direct decays of third--generation squarks into their
lighter partners and Higgs bosons as well as top quark decays into $H^\pm$
bosons. We  show that the production rates of relatively light Higgs bosons,
$M_\Phi \lsim  250$ GeV, via these mechanisms can  be rather large in some
areas of the parameter space. Performing a fast detector simulation analysis 
that takes into account the signals and the various backgrounds, we show that 
the detection of the neutral Higgs bosons through their decays into $b\bar{b}$
pairs, and of the charged Higgs particles through the $\tau^\pm \nu$ signature,
is possible at the LHC.

\end{abstract}

\newpage 

\subsection*{1. Introduction}

The issue of detecting the extended Higgs particle spectrum \cite{HHG} present 
in the Minimal Supersymmetric  Standard Model (MSSM) \cite{MSSM} at
hadron colliders has triggered a large theoretical and experimental activity in
the last twenty years; for recent reviews, see for instance
Ref.~\cite{reviews}. In particular, thorough experimental analyses and
simulations \cite{TDRs} have shown that at least the lightest MSSM CP--even
Higgs particle $h$ should be found at the Large Hadron Collider (LHC). In large
areas of the parameter space that have not been ruled out by LEP2 searches
\cite{LEP2}, the heavier neutral CP--even $H$ boson,  the CP--odd $A$ boson, as
well as the charged Higgs particles $H^\pm$, can also be discovered for an
integrated luminosity as high as $\int {\cal L} \gsim 100$ fb$^{-1}$.  \s

These conclusions were reached by simply investigating direct Higgs boson
production through Standard Model (SM)--like processes\footnote{It is also
assumed, in general, that the Higgs bosons decay only into SM particles and
that loop contributions of SUSY particles do not substantially alter the
rates of the production and decay processes. If these assumptions do not hold,
as might be the case if SUSY particles are relatively light, the searches
could become more complicated; for discussions see for instance
Refs.~\cite{sdecay,sloop}.}: mainly the gluon--gluon fusion mechanism mediated
by heavy--quark loops, $gg \to h,H,A$ \cite{Hgg}, and the associated production
with heavy quarks, $gg/ q\bar{q} \to h,H,A+ b\bar{b}$ or $t\bar{t}$ \cite{Hqq},
for the neutral Higgs particles and top quark decays, $t \to H^+ b$, or
associated production with top quarks $gg/ q\bar{q} \to H^+ b\bar{t}$ and $gb
\to H^-t$ \cite{H+prod}, for the charged Higgs boson. The production cross
sections of most of these processes [those involving $b$ quarks]  are strongly
enhanced for large values of $\tan \beta$, the ratio of the vacuum expectation
values of the two Higgs--doublet fields that break the electroweak symmetry in
the MSSM. \s

Another potential source of MSSM Higgs bosons at the LHC is due to the cascade
decays of squarks and gluinos: because of their strong interactions, these are
copiously produced in hadronic collisions. These particles could then decay
into the heavy chargino and neutralinos $\chi_2^\pm$ and $\chi_{3,4}^0$  and,
 if enough phase space is available, the latter could decay into the lighter
chargino and neutralinos $\chi_1^\pm$ and $\chi_{1,2}^0$  and Higgs
bosons: 
\begin{eqnarray} 
pp \to \tilde{g} \tilde{g} , \tilde{q} \tilde{q}, \tilde{q} \tilde{q}^* ,
\tilde{q} \tilde{g}  & \to & \chi_2^\pm, \chi_3^0, \chi_4^0 + X  \non \\
& \to &   \chi_1^\pm, \chi_2^0, \chi_1^0 + h,H,A, H^\pm \ +X 
\eeq

There is also the possibility of a direct decay of squarks and gluinos into
the lightest charginos $\chi_1^\pm$ and the next--to--lightest neutralinos
$\chi_2^0$ which decay, again if enough phase space is available, into the
lightest neutralino  and Higgs bosons:
\begin{eqnarray} 
pp \to \tilde{g} \tilde{g} , \tilde{q} \tilde{q}, \tilde{q} \tilde{q}^* ,
\tilde{q} \tilde{g}  & \to & \chi_1^\pm, \chi_2^0 + X  \non \\
&\to&   \chi_1^0 + H^\pm, h,H,A \ +X 
\eeq
From now on, we will call the decay chain in eq.~(1) the ``big cascade" while 
the one in eq.~(2) is dubbed the ``little cascade". 
Other possibilities for Higgs particle production are the direct decays of 
heavier top and bottom squarks into the lighter ones and Higgs bosons, if 
large enough squark mass splitting is available \cite{stopH}:
\beq
pp &\to& \tilde{t}_2 \tilde{t}_2^*, \tilde{b}_2 \tilde{b}_2^* \ \ {\rm with} \ \ 
\tilde{t}_2 (\tilde{b}_2) \to \tilde{t}_1 (\tilde{b}_1) +h/H/A \ {\rm or} \ 
\tilde{b}_1 (\tilde{t}_1) +H^\pm 
\eeq
as well as top quarks originating from SUSY particle cascades, decaying into 
$H^\pm$ bosons:
\begin{eqnarray} 
pp \to \tilde{g} \tilde{g} , \tilde{q} \tilde{q}, \tilde{q} \tilde{q}^* ,
\tilde{q} \tilde{g}  & \to & t/\bar{t} + X  \to    H^\pm +X 
\eeq

The production of the lightest $h$ boson from cascade decays of  strongly
interacting SUSY particles into the next--to--lightest neutralinos $\chi_2^0$,
which then decay into the $h$ boson and the lightest neutralinos $\chi_2^0 \to
h \chi_1^0$, has been known for quite some time \cite{hcascade}.  Cascade
decays of squarks and gluinos into relatively light $H,A$ and $H^\pm$ bosons,
where the main contribution is due to the cascades $\chi_1^\pm, \chi_2^0 \to
\chi_1^0+$ Higgs bosons, have also been discussed  in the past; see
Ref.~\cite{oldcascade}. Charged Higgs bosons produced via squark/gluino decays
have recently been considered  \cite{H+cascade} in the range where they cannot
be produced from top decays [i.e. for $M_{H^\pm} \gsim m_t$] or in association
with top quarks [i.e. in the region of moderate $\tb$ values where the $H^+tb$
Yukawa coupling is not enhanced enough].   In the present paper, we extend the
previous analyses in three directions: \s

-- We investigate in detail the cascades of strongly interacting SUSY 
particles into the heavier neutral Higgs particles $H$ and $A$. In particular,
we will analyse the production rates in the situation where these particles
have relatively  small masses, $M_{H,A} \lsim 150$ GeV, and moderate to large
Yukawa couplings to $b$--quarks [the so--called ``intense--coupling regime"
discussed in Ref.~\cite{intense} for instance]. We will also investigate in
more detail  the cascades of strongly interacting SUSY particles into the
lightest $h$ boson, where both types of  cascades, as indicated in eqs.~(1) and
(2), are in general present.  \s

-- In the case of the charged Higgs boson, we extend the analysis of 
Ref.~\cite{H+cascade} to $H^\pm$  masses below the top quark mass, where the
direct decay process $t \to H^+b$ is also at work and acts as an important
source of ``background" events. For the production signal, we will not only
discuss $H^\pm$ bosons  originating from the decays of heavier gauginos, but
also the more complicated situation where they also come from
decays of top quarks originating from the cascade  decays of squarks and
gluinos as in eq.~(4). \s

-- We perform a full Monte Carlo simulation, in which we consider the signals
along with   the various SM and SUSY backgrounds. Using simple kinematical cuts
to reduce the most dangerous backgrounds and, to be as  realistic as possible,
simulating the response of the CMS detector [the conclusions are expected to
remain valid for ATLAS], in particular, the ability to tag $b$ quarks and 
$\tau$ leptons, we show that these processes can indeed be detected in some
regions  of the MSSM parameter space, particularly in areas where there is no 
access to the heavier Higgs bosons in the direct production mechanisms.\s

We will not attempt to make an extremely detailed analysis of this topic. The
subject being too involved and complicated, it calls for rigorous and dedicated
simulation studies to be covered completely. Rather, our  aim will be  to open
the Pandora box and investigate in various  possible scenarios [which are 
hopefully representative of what could broadly happen in the MSSM] this new
source of Higgs bosons at the LHC and show that these particles can indeed be
detected.  There are two main reasons, in our opinion, which motivate such a
study [in addition, of course, to the trivial reason that it is a new source
of MSSM Higgs bosons of unforeseen potential and as such, it must be analysed
anyway]:\s

The first reason is that an important issue at the LHC, once supersymmetric
particles are found, will be the reconstruction of the SUSY Lagrangian at the
low--energy scale, which would allow the structure of the fundamental theory at
high scales to be derived.  This can be achieved only by measuring some of the
couplings between the SUSY and the other particles. Among these, the couplings
of superparticles to Higgs bosons are of special importance, since they also
probe the electroweak symmetry breaking sector and might decide which Higgs
scenario is at work. The cascade decays involve these couplings and would
provide crucial information to achieve this goal. \s

Another argument for this study is that, in the MSSM parameter space, there is
a hole in the region with  130 GeV $\lsim M_A \lsim 170$ GeV and $\tb \sim 5$,
where only the lightest $h$ boson can be found at the LHC [a hole also exists
for $M_A$  larger than $\sim 200$ GeV and $\tb \sim 5$--10]. This is because
the dominant production processes for the heavier neutral Higgs bosons, $gg \to
H/A+ b\bar{b}$ or $t\bar{t}$ and $gg \to H/A$, have not  sufficiently large
cross sections [the Yukawa couplings of the $H$ and $A$  bosons to $b$--quarks,
proportional to $\tb$, are not sufficiently enhanced, while the couplings to
$t$--quarks are suppressed] and the charged Higgs boson  cannot be probed in
top decays since $M_{H^\pm} \gsim m_t-m_b$ and its coupling to  $tb$ pairs [which
is also proportional to $m_b \tb$ and $m_t/\tb$] is not  sufficiently enhanced.
We will show that in the cascade decays, the value of $\tb$ does not play a
crucial role and, therefore, this hole could be filled up by  searches through
these decays. \s

In the next section, we discuss the  production of squarks and gluinos at the
LHC and their decays into the MSSM Higgs bosons through cascades involving
gauginos. We also discuss the  direct decays of top and bottom squarks to the
Higgs bosons and  for $m_t >m_{H^\pm}$, $H^\pm$ production from direct decays
of top quarks coming from $\tilde{q}$ and  $\tilde{g}$ decays. In all cases,
the  cross sections times branching ratios will  be presented in various
scenarios. In section 3, we present our event generator analysis  in selected
areas of the parameter space and show that the signal events can  be detected
in spite of  all the SM and SUSY background processes. A short conclusion will
be presented in section 4. 

\subsection*{2. The signal cross sections} 

At the LHC, the total squark and gluino cross section in the various pair and
associated production processes listed in eqs.~(1) and (2) is of the  order of
$\sigma (\tilde{q}+\tilde{g}) \sim 110$ to 3 pb for sparticle masses 
$m_{\tilde{g}} \sim m_{\tilde{q}} \sim 0.5$ to 1 TeV, leading  to a large, $
\sim 3 \times 10^{7}$ to $10^{6}$, number of events with an accumulated
luminosity of $\int {\cal L} \sim 300$ fb$^{-1}$.  These squarks and gluinos
can decay into the heavier chargino and neutralinos $\chi_2^\pm, \chi_3^0$ and
$\chi_4^0$, with large branching fractions of about a few times ten per
cent. If enough phase space is available, the latter particles could then decay
into the lighter chargino and neutralinos, $\chi_1^\pm, \chi_1^0$ and
$\chi_2^0$, and neutral $h,A,H$ or charged $H^\pm$ bosons, with branching
ratios again of the order of a few times ten per cent, making a total of a few percent
branching ratio for the whole cascade. If the mass splitting between the
lightest chargino or next--to--lightest neutralino and the LSP neutralino
is large enough, another source of Higgs bosons will appear in a more direct way
from the decays of $\chi_1^\pm, \chi_2^0 \to \chi_1^0 +$ Higgs bosons, which
has a rather large branching ratio.  In some situations both possibilities are
at work, leading to a substantial number of Higgs particles in the final state.
[In the case of a relatively light charged  Higgs boson, $M_{H^\pm} \lsim m_t$,
there is an additional production  mechanism, $t \to bH^+$, with the
$t$--quarks produced either directly or from the cascade decays of squarks and
gluinos.]\s

A key point in this analysis, is that the coupling of the Higgs bosons to
chargino and neutralino states is maximal for higgsino--gaugino mixed states
\cite{Haber}, while the gauge boson couplings to neutralinos are maximal for
higgsino--like states. In the gaugino--like [i.e.\ when the higgsino mass
parameter $|\mu|$ is much larger than the gaugino mass parameter $M_2$] or
higgsino--like [i.e. in the opposite situation $|\mu| \ll M_2$] regions, this
results into the dominance of the decays of the heavier charginos and
neutralinos into the lighter ones and Higgs bosons, over  the same decays
with gauge boson final states in general. This is also the case of the little
cascades in the gaugino region;  the branching ratios for the decays of
$\chi_2^0$ and $\chi_1^+$ into the LSP and Higgs bosons, when kinematically
accessible, are in general more important than for gauge boson final states.
\s 

We will closely follow the analysis of Ref.~\cite{H+cascade}, where the cross
sections for squark and gluino production at the LHC have been discussed and
where their decay branching ratios into the heavier charginos and neutralinos,
which then decay into the lighter ones and charged Higgs bosons, have been
analysed in detail. The analytical expressions of the various partial decay
widths, including also the decays into neutral Higgs particles and the little
cascades, have been given there.  Here, we will simply show the total
production cross sections times the decay branching ratios for the final states
involving {\it a single} $h,H, A$ and $H^\pm$ particle in various scenarios.
Throughout the analysis, we will assume the universality of the 
soft--SUSY--breaking gaugino mass parameters at the high--energy scale, leading
to the relation $m_{\tilde{g}} \simeq M_3 \sim 3M_2 \sim  6M_1$ at low
energies. \newpage

\subsubsection*{2.1 The various scenarios}

We will discuss the following four scenarios, which cover most of the
possibilities for the cascade decays that we will discuss and span over the
typical MSSM parameter space. \s

{\bf -- Scenario Sc1:} Here, the scalar partners of light quarks are heavier 
than gluinos [and top squarks].  For illustration, we chose
$m_{\tilde{q}} =1.2 m_{\tilde{g}}=720$ GeV with the gluino mass taken to be
$m_{\tilde{g}} =600$ GeV. The first-- and second--generation squarks will then
decay dominantly into quarks and gluinos, $\tilde{q} \to \tilde{g}q$ [a 
smaller fraction will decay into quarks and chargino or neutralino final
states]. In the case of lighter top squarks, since their mass is smaller than
$m_{\tilde{g}}+m_t$, they will directly decay into chargino+bottom and
neutralinos+top final states. The gluinos, which are directly produced or which
come from the decays of squarks, will mainly undergo three--body decays into 
$q \tilde{q} \chi_i^0$ and $q \tilde{q}^\prime \chi_i^\pm$.  This decay is
dominantly mediated by the exchange of the lightest top squarks, which have a
smaller virtuality [i.e. a mass smaller than the common scalar quark 
mass], thanks to the larger mixing between the stop eigenstates, which is
proportional to $m_t$. In this scenario, we will vary the higgsino mass
parameter $\mu$ while the bino and wino mass parameters $M_1$ and $M_2$ are
fixed to $M_2=200$ GeV and $M_1=100$ GeV by our choice of $m_{\tilde{g}}$, 
by virtue of the assumption of gaugino mass unification.\s

{\bf -- Scenario Sc2}: This scenario is similar to the previous one, i.e.  we
assume that squarks are heavier than gluinos with the same ratio of 
$m_{\tilde{q}} =1.2  m_{\tilde{g}}$,  but there is a major difference:  here we
take the gluino mass to be larger,  $m_{\tilde{g}} =900$ GeV, which leads to
$M_2=2M_1=300$ GeV. For large enough $\mu$ values, the lighter charginos and 
neutralinos are gaugino--like with masses $m_{\chi_1^\pm} \sim m_{\chi_2^0} 
\sim 2 m_{\chi_1^0} \sim M_2$. Thus for $M_A \lsim 130$ GeV, the Higgs bosons
are light enough, to render the mass splittings among the lighter gaugino states
such that  the decays $\chi_1^\pm \to \chi_1^0 H^\pm$ and $\chi_2^0 \to 
\chi_1^0  h,H$ and $A$ can occur. Therefore, both the big cascade, which is the
only one present in Sc1,  and the small cascade of eq.~(2) are at work
together in this scenario. \s

{\bf -- Scenario Sc3:} Here gluinos are heavier than squarks, and we choose the
common squark mass to be $m_{\tilde{q}}= 800$ GeV. Therefore, the decays
$\tilde{g} \to \tilde{q} q$ occur 100\% of the time. Virtually, the 
electroweak cascades start with only the squark states, those coming from 
direct production and those from gluino decays. Here, we fix $\mu=150$ GeV and
hence we are in the higgsino--like region, i.e.  with a small value of the
$\mu$ parameter with respect to $M_2$ [again universal gaugino masses are
assumed at the high scale].  In this case, all squarks [in particular those of
the first two generations whose couplings to the higgsino-like $\chi_1^\pm,
\chi_1^0, \chi_2^0$ states  are proportional to the small mass of their partner
quarks and are therefore tiny] will mainly decay into the heavier chargino and
neutralinos [which are gaugino--like with masses $m_{\chi_2^+} \sim
m_{\chi_4^0} \sim 2 m_{\chi_3^0} \sim M_2$]. For large enough $M_2$ values,
there is sufficient phase space for the decay of the heavier gauginos into the
lighter higgsino states, with masses $m_{\chi_1^+} \sim m_{\chi_1^0} \sim
m_{\chi_2^0} \sim |\mu|$, and Higgs particles with masses $M_\Phi \lsim 200$
GeV to occur.  \s

{\bf -- Scenario Sc4:} This scenario is the same as the previous one as far as
the squark and gluino sectors are concerned, but it is different for the
electroweak gaugino sector. Indeed, here we choose a large value for the
higgsino mass parameter, $\mu=1$ TeV, which now makes the lighter chargino and
neutralinos gaugino--like. Squarks and gluinos will then decay into these
states and, for large enough values of $M_2$, the lightest chargino and the
next--to--lightest neutralino can decay into the LSP and Higgs bosons [contrary
to Sc3, where the little cascade was absent since $\chi_1^\pm,
\chi_{1,2}^0$ where higgsino--like and degenerate in mass]. Of course, in this
scenario, the big cascade is in principle kinematically possible but the
branching ratios of first-- and second--generation squark decays into higgsino
states are strongly suppressed because of the small squark--quark--higgsino
coupling. \s

The Higgs sector will be treated using the program {\tt HDECAY}, version 2.0
\cite{hdecay}, which includes the important radiative corrections. The masses
of the neutral $h,H$ and charged $H^\pm$ particles are obtained once the two
input parameters at the tree--level, $M_A$ and $\tb$ [as well as the scalar
masses and other soft--SUSY--breaking parameters $\mu, A_b$ and $A_t$, which
enter the radiative corrections] are fixed. We will choose for illustration the
input values: $M_A=100, 130$ and 200 GeV and $\tb=10,30$, which leads to the
CP--even and charged Higgs boson masses given in Table 1. We will also discuss
the scenario with $M_A=150$ GeV and $\tb=5$, where at the LHC only the $h$
particle can be discovered in the direct production mechanisms, as discussed
previously; the masses of the other Higgs bosons are also given in Table 1. \s

To evade the LEP2 bounds on the light $h$ boson mass\footnote{The absolute
experimental bound from LEP2 searches of the $h$ boson mass is $M_h \gsim 92$ 
GeV, unless the Higgs boson is SM--like; in this case, the bound becomes $M_h \gsim 114$
GeV \cite{LEP2}. This rules out  small values of the parameter $\tb$ in the
absence of mixing in the stop sector.} for small values of $M_A$ and/or $\tb$,
we use a large value for the trilinear couplings $A_t \gsim m_{\tilde{q}}$, 
which increases the value of $M_h$ [the so--called typical--mixing scenario].
This leads to a large splitting in the stop sector: for $m_{\tilde{q}}=960$ GeV
and $A_t=1.5$ TeV for instance, one has $m_{\tilde{t}_1} \sim 840$ GeV and
$m_{\tilde{t}_2} \sim 1.1$ TeV [with a small variation with $\mu$ and $\tb$],
which also favours the cascade decays in Sc1 and Sc2, since the
virtuality of the lighter $\tilde{t}_1$ is smaller than that of the other
squarks, enhancing the three--body decays of gluinos into higgsino final
states. \s

\begin{table}[htbp]
\renewcommand{\arraystretch}{1.45}
\begin{center}
\vspace*{-3mm}
\begin{tabular}{|c|c|c|c|c|} \hline
$M_A$ [GeV] & $ \ \ \tb \ \ $ & $M_h$ [GeV] & $M_H$ [GeV] & $M_{H^\pm}$ [GeV] 
\\ \hline
100 & 10 & 97 & 125 & 127 \\
    & 30 & 99 & 123 & 126 \\ \hline
130 & 10 & 116 & 136 & 151 \\
    & 30 & 121 & 130 & 151 \\ \hline
200 & 10 & 120 & 201 & 215 \\
    & 30 & 122 & 199 & 214 \\ \hline \hline
150 & 5 & 111 & 160 & 169 \\ \hline
\end{tabular}
\end{center}
\vspace*{-3mm}
\caption[]{\it Masses of the neutral and charged Higgs bosons for different 
inputs for $M_A$ and $tan \beta$ [$m_{\tilde{q}}=960$ GeV, $A_t=1.5$ TeV, 
$\mu=100$ GeV and $M_2=270$ GeV].}
\vspace*{-3mm}
\end{table}

The produced neutral Higgs bosons will decay mainly into $b\bar{b}$ and $\tau^+
\tau^-$ pairs, with branching ratios of respectively 
$\sim 90\%$ and $\sim 10\% 
$,
except for the CP--even $h$ and $H$ bosons when they are almost SM--like
[for high values of $\tb$ this occurs already when $M_A \gsim (\lsim) 130$ GeV 
for the $h\, (H)$ states], where additional channels [such as $WW^{(*)}$
decays]  occur and will slightly suppress these rates.  Depending on whether
the $tb$  threshold is reached or not, the $H^+$ bosons decay mainly into $tb$
or  $\tau^+ \nu$ final states.  \s

The cross sections for squark and gluino production are evaluated as in Ref.\,
\cite{H+cascade}, using the CTEQ3L \cite{CTEQ} parametrization of the parton
densities and with the scale defined as the average of the masses of the final
sparticles. To be conservative, we will neither include the $K$--factors, which
enhance the production cross sections \cite{Kfactors}, nor the possibly large
QCD corrections to squark and gluino decays \cite{QCDdec}. For the decay
branching ratios, we use the average one defined in
Ref.~\cite{H+cascade} for squark decays when they are heavier than gluinos
[Sc1 and Sc2], i.e.  that we sum over all possibilities for decaying
left-- and right--handed  squarks as well as for up-- and down--type squarks,
keeping track of flavour and chirality to obtain a specific final state. For
stops, we take into account the direct decays of the heavier ones into the
lighter ones and Higgs or gauge bosons, but not the higher order decay modes
\cite{Dtops}. For the three--body decays of gluinos, we take all possible
channels into account and include masses for the third--generation fermions and
sfermions and full sfermion mixing \cite{Dgluino}.\s

For charged Higgs bosons lighter than the top quark, we will also discuss their
production from  top decays. For moderate values of  $\tb$, $5 \lsim \tb \lsim 
20$, the branching fraction of the decay $t \to bH^+$ is rather tiny and the 
number of $H^\pm$ bosons from the SUSY cascade decays is  small.  For larger
$\tb$ values, this branching fraction can be rather  large [a few times ten 
per cent] and would lead to a substantial number of $H^\pm$  bosons from SUSY
cascades. The top quarks are either produced directly, $pp \to gg/q\bar{q} \to
t\bar{t}$ or come from cascade decays of squarks and gluinos: two--body,  $g
\to  t \tilde{t}_{1,2}^*, b \tilde{b}_{1,2}^*$,  or three--body, $\tilde{g} \to
t\bar{t} \chi_i^0, t\bar{b}  \chi_i^+$, decays of gluinos  and two--body decays
of stops and sbottoms in top quarks and $\chi$ states, $\tilde{t}_i \to t
\chi_j^0$ and  $\tilde{b}_i \to t \chi_j^-$.  

\subsubsection*{2.2 Cascades involving gauginos}

The variation of cross sections times branching ratios to obtain at least {\it
one} neutral or  charged MSSM Higgs boson in the final state from the big or
little cascade, or from both, is shown as a function of $\mu$  in Figs.~1 and
2 for Sc1 and Sc2, respectively, and as a function of $M_2$ in Figs.~3 and 4
for Sc3 and Sc4. For each set, we take $M_A=100, 130$ and  200 GeV for the 
top, middle and bottom rows respectively, and $\tb=10$ and 30 for left and
right panels, respectively. Fig.~5 shows the rates for a specific choice of
$M_A=150$ GeV and $\tb=5$ in the above four scenarios.\s

Let us first make a few general  remarks on these figures.
In all of them, we see that $\sigma \times$BR for the four Higgs bosons can be
rather large, exceeding the level of 0.1 pb in very large portions of the
parameter space, and even reaching the level of $\sim 10$ pb  in some cases. 
This means that more than $3 \times 10^{4}$ events, and up to $3 \times 10^{6}$
events, can be collected at the LHC for an integrated luminosity of $\int {\cal
L} \sim 300$ fb $^{-1}$. In the cases of $A,H$ and $H^\pm$ bosons, the
production rates are larger for smaller values of the masses [as expected from
simple phase--space considerations], while they are stable for $h$ boson
production [since the variation of $M_h$ is mild for $M_A=$100--200 GeV].  The
dependence of the rates on the value of $\tb$ is not very pronounced in
general, in contrast to the very strong $\tb$--dependence  for Higgs production
in  standard--like processes such as gluon--gluon fusion or associated 
production with $b\bar{b}$ pairs. \bigskip

{\bf In Sc1} with $M_3=3M_2=6M_1=600$ GeV, the mass differences between the
lighter $\chi_1^\pm$/$\chi_2^0$ states and the $\chi_1^0$ LSP are less than
the minimum value of the lightest $h$ boson mass [and thus also the $A,H$ and
$H^\pm$ boson masses] considered, $M_h \gsim 100$ GeV, and therefore the
little cascades, $\chi_2^0 \to h,H,A+\chi_1^0$ and $\chi_1^\pm \to
H^\pm+\chi_1^0$, are kinematically not allowed. The variation of $\sigma
\times$BR with $\mu$ is simply due to the variation of the branching ratios BR$(\tilde{g}
\to \chi_{3,4}^0 qq, \chi_2^\pm qq')$ and  BR$(\chi_{3,4}^0, \chi_2^\pm \to
\chi_{1}^\pm, \chi_{1,2 }^0$ + Higgs), since the cross sections for squark and
gluino production are constant, $m_{\tilde{g}}$ and $m_{\tilde{q}}$ being
fixed. The production rates are larger for small (or moderate) $\mu$ values,
$\mu \lsim 400$ GeV, where the gaugino (higgsino)--like states $\chi_2^\pm$ and
$\chi_{3,4}^0$ have enough phase space to decay into the higgsino
(gaugino)--like states $\chi_1^\pm$ and $\chi_{1,2}^0$. For larger values of
$\mu$, BR$(\tilde{g} \to \chi_{3,4}^0 qq, \chi_2^\pm qq')$ become smaller
because of phase--space suppression, and $\sigma \times$BR drops out. \s

For $M_A \lsim 150$ GeV, the signal drops sharply for the CP--even and charged
Higgs bosons for intermediate values of $\mu \sim M_2 \sim $ 200--250 GeV. In
this case, the charginos and neutralinos are mixtures of gauginos and higgsinos
and their mass differences are rather small, leading to phase--space--suppressed
decays of the heavier states into the lighter ones and Higgs bosons.  This is
more pronounced for $H^\pm$ [and also $H$], which is the heaviest Higgs boson. 
However, the production rate for the CP--odd $A$ boson is enhanced for small
$M_A$ values.  This is due to a conjunction of several facts: the $A$ boson is,
together with $h$, the lightest Higgs particle and is therefore more favoured by
phase space\footnote{In particular, the decay channel $\chi_3^0 \to
\chi_1^0$+Higgs can be kinematically open for the $h$ and $A$ bosons and closed
for the $H$ and $H^\pm$ particles. Note also that the  phase--space suppression
is different for decays of charginos and neutralinos into CP--even and CP--odd
Higgs bosons.}; its couplings to charginos and neutralinos are stronger than
those of the $h$ boson. Since the decays to Higgs particles share the branching
ratio [together with the decays into gauge bosons], the suppression or the
absence of decays into other heavier Higgs particles leads to an effective
enhancement of the branching ratio for the decays into the pseudoscalar $A$
boson. \s

Note that for larger Higgs mass values, $M_A \gsim 150$ GeV [say, $M_A=200$
GeV], there is not enough phase space for decays into the heavier $H,A,H^\pm$
bosons in the small and intermediate $\mu$ range, $\mu \lsim 250$, GeV and only
decays into the lighter $h$ boson are allowed. For larger $\mu$ values, $\sigma
\times$BR follows the same trend for $H,A$ and $H^\pm$ production, and even for
$h$ production for very large $\mu$ values [for $H$ and $A$, this is expected
since we are  almost in the decoupling regime where they  have the same masses
and couplings]. The kinks followed by humps in the higher $\mu$ side are due to
the opening of new decays channels, in particular channels involving the
neutralino $\chi_3^0$, which is lighter than $\chi_2^\pm$ and
$\chi_4^0$.\bigskip

{\bf In Sc2}, the cross sections times branching ratios for the four Higgs
particles are smaller than in the previous case for low to medium values of the
$\mu$ parameter, $\mu \lsim 300$ GeV, where only the big cascades are
kinematically allowed. This is due to the larger gluino and squark masses
considered, $m_{\tilde{q}}=1.2 m_{\tilde{g}}=1080$ GeV with $m_{\tilde{g}}=900$
GeV, which make the production cross sections of strongly interacting particles
smaller.  Nevertheless, $\sigma \times $BR is still rather large, being in
general between 0.1 pb and a few picobarns for $M_A \lsim 150$ GeV [except in
the case of $H$ production, because of the smaller couplings to chargino and
neutralino states] and the trend follows the one of the previous scenario.\s

For larger values of the higgsino mass parameter, $\mu \gsim 300$ GeV, and
since in this scenario $M_2=2M_1=m_{\tilde{g}}/3=300$ GeV, we gradually enter
the gaugino region where $m_{\chi_1^\pm}=m_{\chi_2^0}=2 m_{\chi_1^0}$.
Hence,    there is now enough phase space for the lightest chargino and the
next--to--lightest neutralino to decay into the LSP and a Higgs boson, at least
for small $M_A$ values. For $M_A=100$ GeV, the little cascade occurs for all
four Higgs bosons, but $\sigma \times$BR is much larger for $h$ and $H^\pm$
production [where it reaches the level of a few picobarns] than for the
production of $A$ and $H$, in particular for $\tb \sim 10$. This is mainly due
to the strength of the chargino--neutralino couplings to Higgs bosons as
discussed previously. For $M_A \simeq 130$ GeV, only the little cascades into
neutral Higgs bosons are kinematically allowed, while for $M_A \gsim 150$, only
the little cascade into the $h$ boson is kinematically allowed, the other Higgs
particles being too heavy. \newpage

{\bf In Sc3}, the cross sections times branching ratios for the
production  of the four Higgs particles from SUSY cascades are displayed as
functions of $M_2$, with the higgsino parameter fixed to $\mu=150$ GeV. The
common squark mass is chosen to be $m_{\tilde{q}}=800$ GeV while the gluino
mass varies with the gaugino mass parameter as $m_{\tilde{g}}=3M_2$. The values
of $\sigma \times$BR are rather large, being between 0.1 pb and a few
picobarns, as soon as $M_2$ exceeds the level of $\sim 250$ GeV. In this case
the decay channels of the heavier chargino and neutralinos [which are
gaugino--like  with masses $m_{\chi_2^\pm} \sim m_{\chi_4^0} \sim M_2$], into
the lighter higgsino--like ones [with masses $m_{\chi_1^\pm} \sim
m_{\chi_{1,2}^0} \sim |\mu|$], are open only when $|M_2  - \mu| \gsim
M_{\Phi}$. This leads to a very large sample of signal events, $\sim 10^5$ with
the expected luminosity of 300 fb$^{-1}$ at the LHC. For larger $M_2$ values,
the decay of the neutralino $\chi_3^0$, with a mass $m_{\chi_3^0} \simeq M_1
\simeq \frac{1}{2}M_2$ also opens up. For lower $M_2$ values, the mass
splitting between the heavier and lighter chargino and neutralino states is too
small, and the decays into Higgs bosons are not kinematically possible. Note
that because the lighter chargino and neutralinos are almost degenerate in
mass, the little  cascades are absent. \s

The production rates are larger for the $h$ and $H^\pm$ bosons: for $h$ it is
in general due to its lightness and consequently to having a more favourable
phase--space, while for $H^\pm$ it is due to the fact that there are more
possibilities for charged decays to occur [$\chi_{3,4}^0 \to \chi_1^\pm H^\mp, 
\chi_2^\pm \to \chi_{1,2}^0 H^\pm$] and that charged current couplings are
stronger than neutral current couplings.  The production rates are relatively
smaller for $H$ and $A$ boson production and we notice the equal production
rate for large $M_A$ values, when we are close to the decoupling limit. Note
also that the production rates decrease with  increasing $M_2$ since, according
to our assumption of universal gaugino masses at the high scale,
$m_{\tilde{g}}$ increases and the cross section for  gluino production drops
out in the first place.\bigskip

{\bf In Sc4}, the higgsino mass parameter is rather large: $\mu =1$
TeV. This  leads to very heavy $\chi_{3,4}^0$ and $\chi_2^\pm$ [higgsino]
states.   The big cascades are disfavoured in this region for two reasons.
First, squarks being relatively lighter [$m_{\tilde q}=800$ GeV], they cannot
decay into  these heavy higgsinos. Secondly, three--body decays of heavy
gluinos into the heavier  higgsinos are suppressed by an extra factor of
electroweak coupling--squared  [compared to a two-body decay]  and the gluinos
mainly decay into squarks and quarks. Only the little cascades $\chi_2^0\to
h,H,A+\chi_1^0$ and $\chi_1^\pm \to H^\pm +\chi_1^0$ are possible if the Higgs
bosons are  light enough [the cascades are always possible for the production
of the  lighter $h$ boson]. When these cascades occur, the cross sections
times  branching ratios are rather large, exceeding the level of 0.1 pb for not
too large  $M_2$ values and reaching the level of $\sim 10$ pb for light Higgs
bosons and  small $M_2$ values. For larger values of the gaugino mass
parameter, $\sigma  \times$BR drops following the decreasing phase space in the
decay $\tilde{q} \to \chi_2^0 q, \chi_1^\pm q'$, while a further dampening is
caused by the lower gluino production rate due to a simultaneous increase in
gluino mass. \s 

\begin{figure}[htbp]
\begin{center}
\vspace*{-2.2cm}
\centerline{\epsfig{file=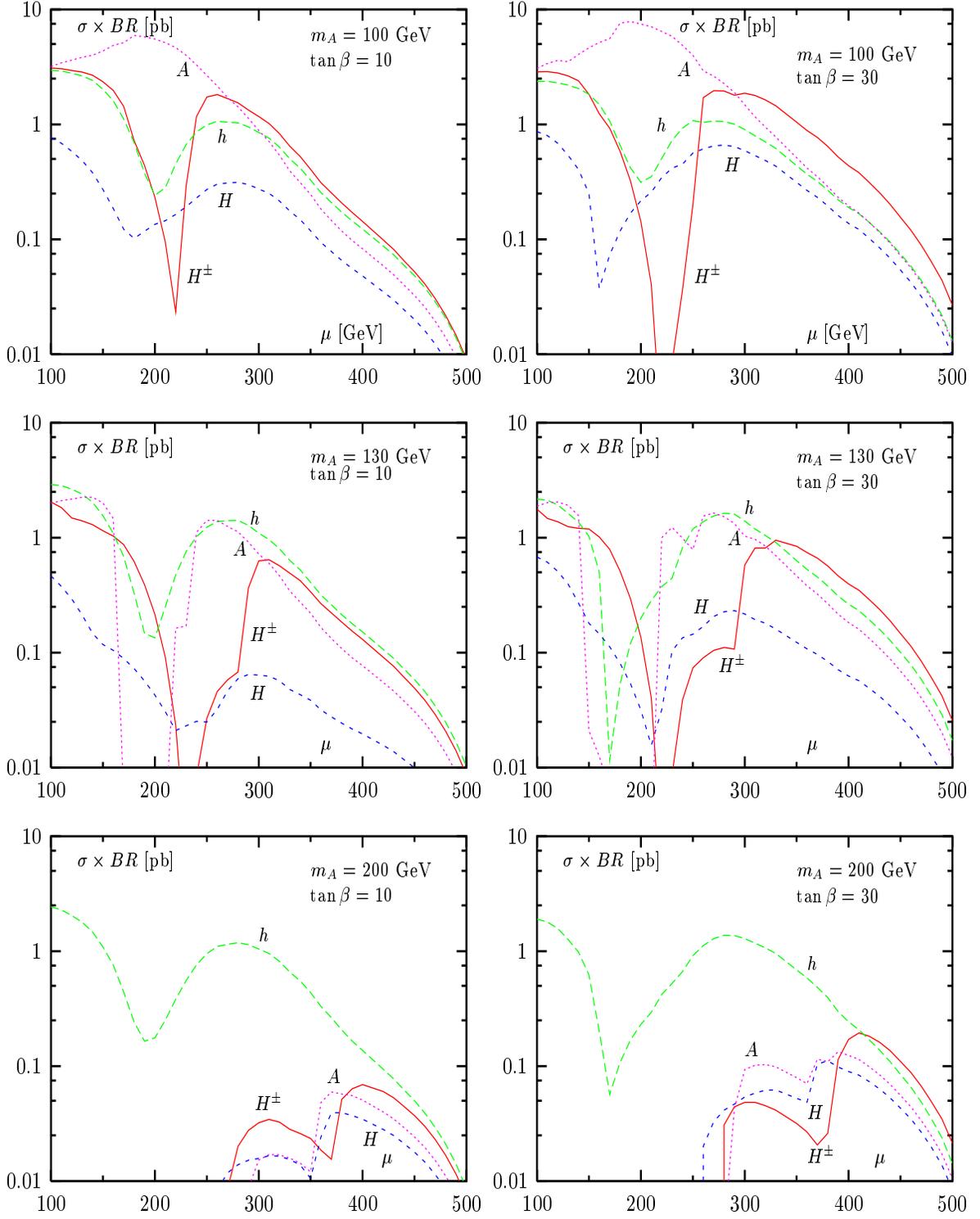,width=19cm}}
\vspace*{-4.5cm}
\caption{\it Cross sections times branching ratios for Higgs production 
in Sc1.}
\end{center}
\end{figure}

\begin{figure}[htbp]
\begin{center}
\vspace*{-2.2cm}
\centerline{\epsfig{file=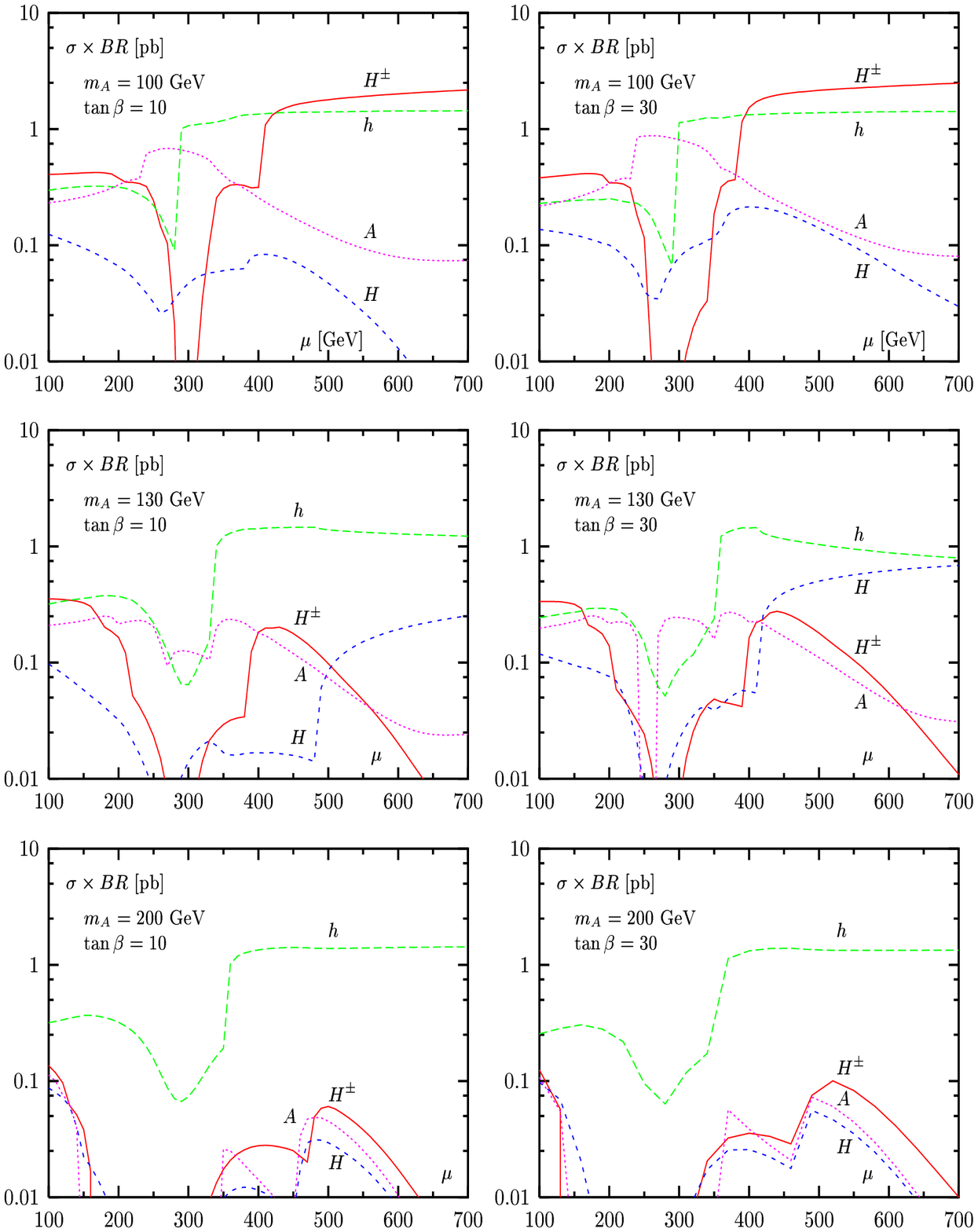,width=19cm}}
\vspace*{-4.5cm}
\caption{\it Cross sections times branching ratios for Higgs production in 
Sc2.}
\end{center}
\end{figure}

\begin{figure}[htbp]
\begin{center}
\vspace*{-2.2cm}
\centerline{\epsfig{file=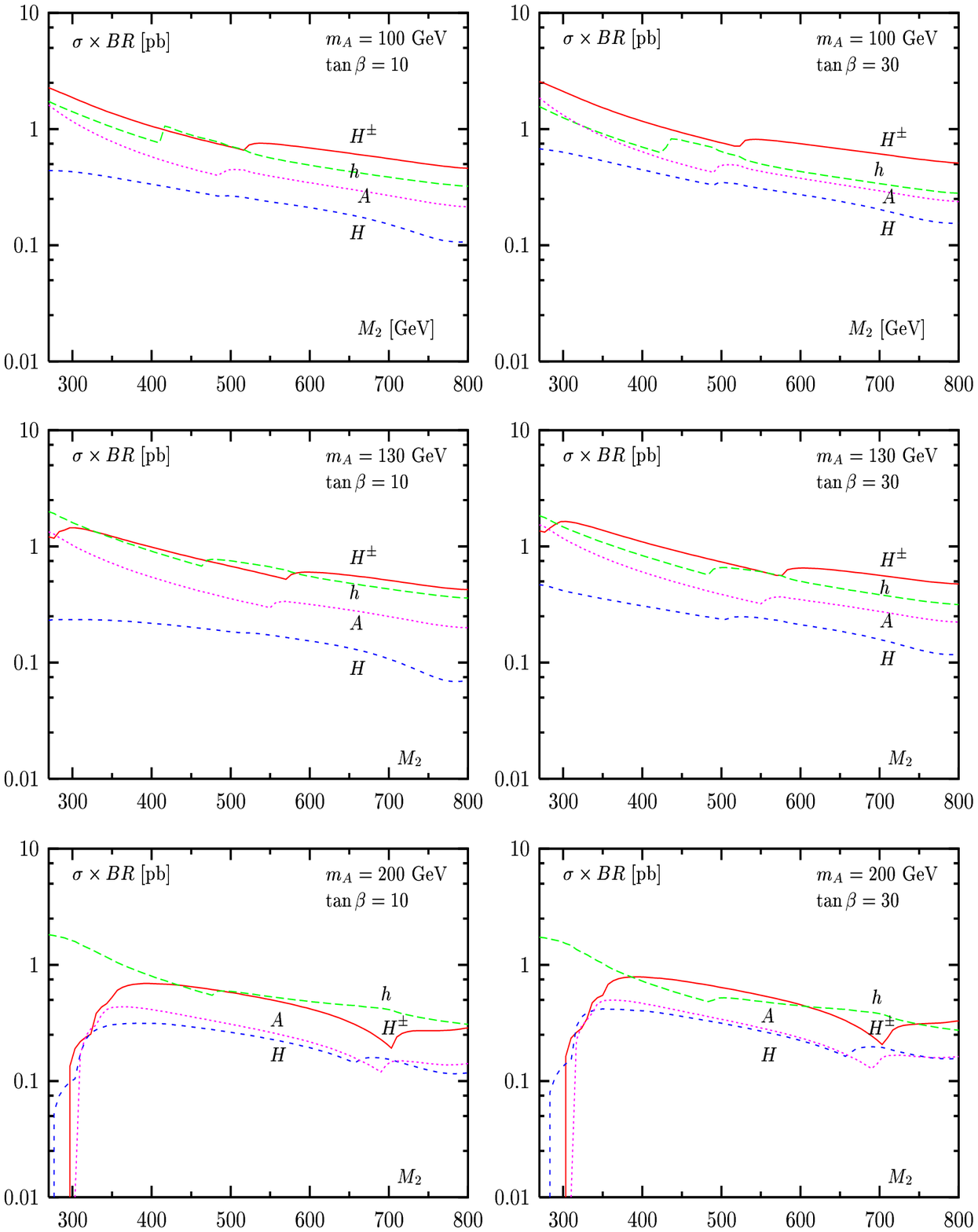,width=19cm}}
\vspace*{-4.5cm}
\caption{\it Cross sections times branching ratios for Higgs production in 
Sc3.}
\end{center}
\end{figure}

\begin{figure}[htbp]
\vspace*{-2.2cm}
\begin{center}
\centerline{\epsfig{file=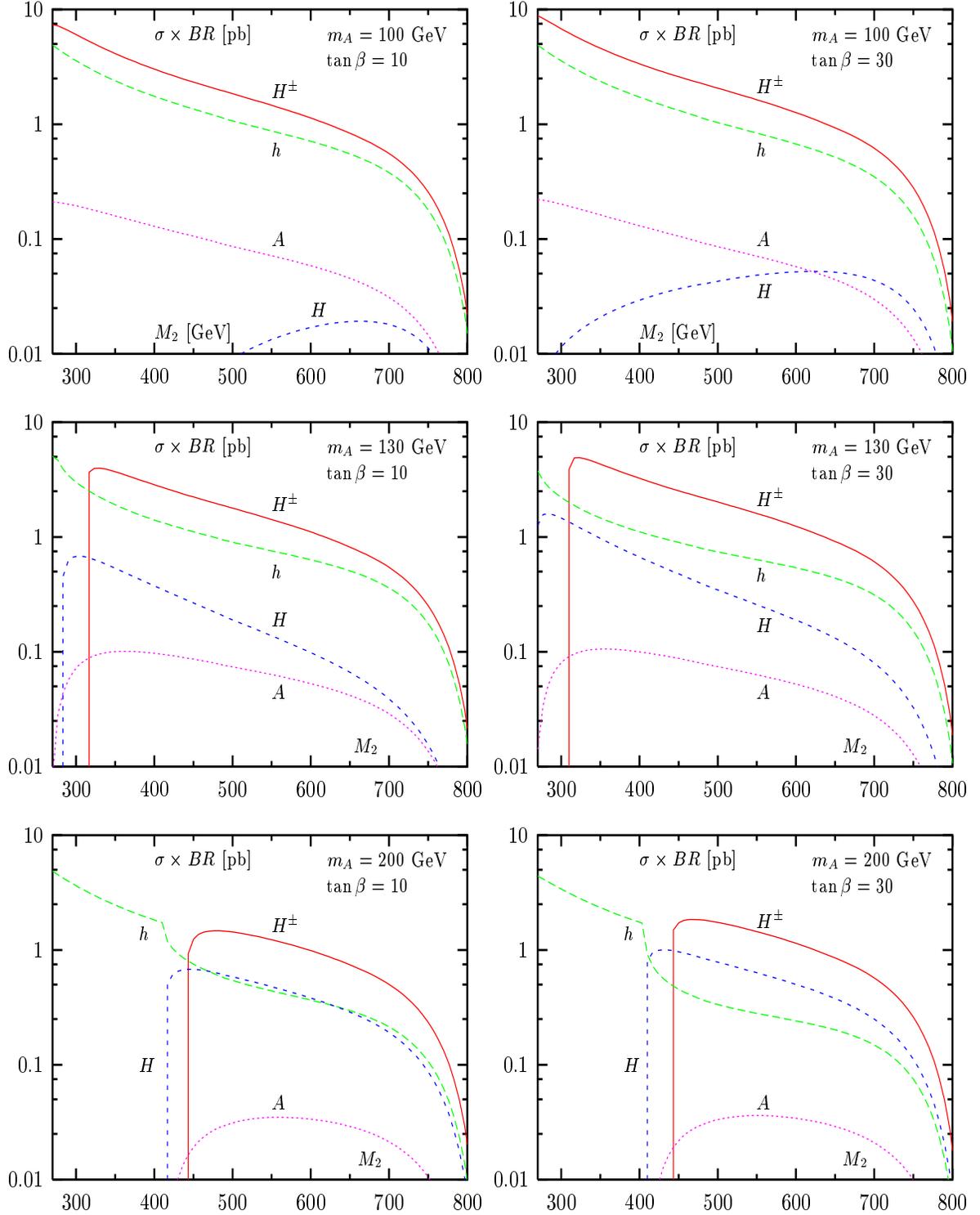,width=19cm}}
\vspace*{-4.5cm}
\caption{\it Cross sections times branching ratios for Higgs production 
in Sc4.}
\end{center}
\end{figure}

Finally, in Fig.~5, we show the cross sections times branching ratios in the
four selected scenarios in the case $M_A=150$ GeV and $\tb=5$. Since, as we
discussed earlier, the dependence on the value of $\tb$ is not very 
pronounced, the trend is half--way between what occurs for $M_A=130$ GeV and for
$M_A=200$ GeV. Except for the regions of intermediate $\mu$ value in 
Sc1 and Sc2 and small $M_2$ values in Sc4, $\sigma \times$ BR for the
production of  the heavier Higgs bosons $H,A$ and $H^\pm$ via the cascades are
substantial, exceeding the level of 0.1 fb in large regions and even reaching
the level of a few picobarns in some cases.

\begin{figure}[htbp]
\vspace*{-2.9cm}
%\vspace*{-6.5cm}
\begin{center}
%\centerline{\epsfig{file=complete_filip-new.ps,width=19cm}}
\centerline{\epsfig{file=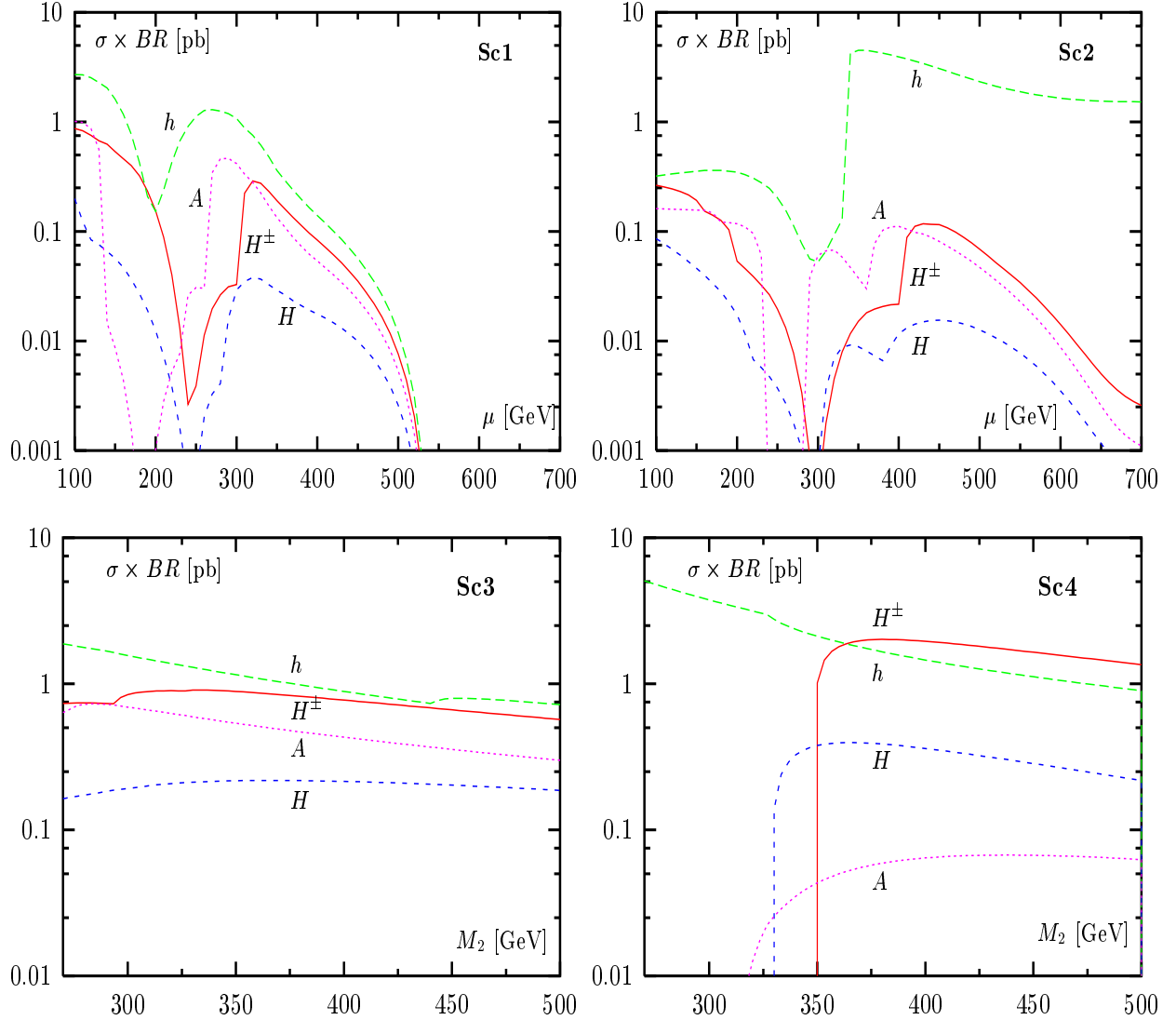,height=30cm,width=20cm}}
\vspace*{-12.cm}
%\vspace*{-3.3cm}
\caption{\it Cross sections times branching ratios for MSSM Higgs boson 
production for $M_A=150$ GeV and $tan \beta=5$ in the four scenarios Sc1 to Sc4.}
\end{center}
\vspace*{-1cm}
\end{figure}

\newpage

\subsubsection*{2.3 Higgs bosons from direct decays of stops and  sbottoms}

If the mass splitting between two squarks of the same generation is large
enough, as is generally the case of the $(\tilde{t},\tilde{b})$ iso--doublet,
the heavier squark can decay into a lighter one plus a Higgs boson $\Phi=h,H,A$
[or a $Z$ boson]. If, in addition, there is enough mass splitting between the
stops and sbottoms, it is possible for the heavier one to decay into the
lighter one and $H^\pm$ [or $W^\pm$] states. The case of charged Higgs bosons has been
studied in detail in Ref.~\cite{H+cascade} and we concentrate here on the stop
and sbottom decays into the neutral Higgs particles, $\tilde{t}_2  \to
\tilde{t}_1 \Phi$ and $\tilde{b}_2 \to \tilde{b}_1 \Phi$. \s

While the production mechanisms for stops and sbottoms are mainly governed by
QCD and the cross sections depend only on the mass of these particles [as well
as on the gluino mass when the decays $\tilde{g} \to \tilde{t}t, \tilde{b}b$
are kinematically possible], their subsequent decay rates  into Higgs particles
depend on many parameters. First, since these squarks are heavy, they have many
possible decay modes besides the ones into Higgs bosons discussed here: decays 
into charginos/neutralinos and quarks, which in general have rather sizable 
decay widths, as well as decays into gauge bosons. In addition, these decays 
depend strongly on the Higgs--squark couplings, which in the case of the
neutral Higgs particles are given by [see Ref.~\cite{H+cascade} for a
discussion of  the couplings and decay widths]:  
\beq  
g_{\tilde{q}_1 \tilde{q}_2 h} &\propto & M_Z^2 \sin2\theta_q (2I_q^3-4 e_q 
s_W^2) \sin (\alpha+\beta)  +2 m_q \cos2\theta_q (A_q r_2^q + 2I_q^3 \, \mu \, 
r_1^q) \non \\  
g_{\tilde{q}_1 \tilde{q}_2 H}& \propto &   - M_Z^2 \sin 2\theta_q (2I_q^3-4 
e_q s_W^2) \cos (\alpha+\beta)  +2 m_q \cos2\theta_q (A_q r_1^q - 2I_q^3 \, 
\mu \, r_2^q)  \non \\  
g_{\tilde{q}_1 \tilde{q}_2 A} &=& - g_{\tilde{q}_2 \tilde{q}_1 A} \propto  m_q
\bigg[ \mu +A_q (\tan \beta)^{-2I_q^3} \bigg]   
\eeq  
where $\alpha$ is the mixing angle in the CP--even Higgs sector of the MSSM, 
$\theta_q$ is the squark mixing angle [for $\theta_q=0$, $\tilde{q}_1=
\tilde{q}_R$ in general], and the coefficients $r^q_{1,2}$ are the
normalization factors of the $h,H$ boson couplings to fermions normalised to the
SM Higgs boson coupling. The largest component of the CP--even Higgs boson
couplings is the one proportional to the quark mass [in the case of sbottom
couplings, a $\tb$ enhancement is present for the coupling of the $H$ state]. 
However, if the mass splitting between the two squarks is large, the squark 
mixing angle approaches the value $|\theta_q| \sim \pi/4$ and the term $\cos
2\theta_q$ approaches zero [the other component of the coupling, the one
proportional to $M_Z$, although maximal since $\sin2\theta_q \sim 1$, is not
large enough to make the Higgs--squark coupling strong]. In the case of no
mixing, where the coupling becomes strong, there is in general not  large
enough mass splitting between the two squarks [at least if one assumes common 
scalar mass terms] to make the decays into Higgs bosons possible. Thus, the
only decays that could have sizeable partial widths would be the channels  
$\tilde{t}_2 (\tilde{b}_2) \to \tilde{t_1}(\tilde{b}_1)A$ whereas, in the case
of  sbottoms, a large value of $\tb$ is needed to make the sbottom mass
splitting  large for the decay to occur and to compensate for the smallness of
$m_b$ in  the coupling to have a sizeable rate. \s

\begin{figure}[htbp]
\begin{center}
\vspace*{-1.2cm}
\centerline{\epsfig{file=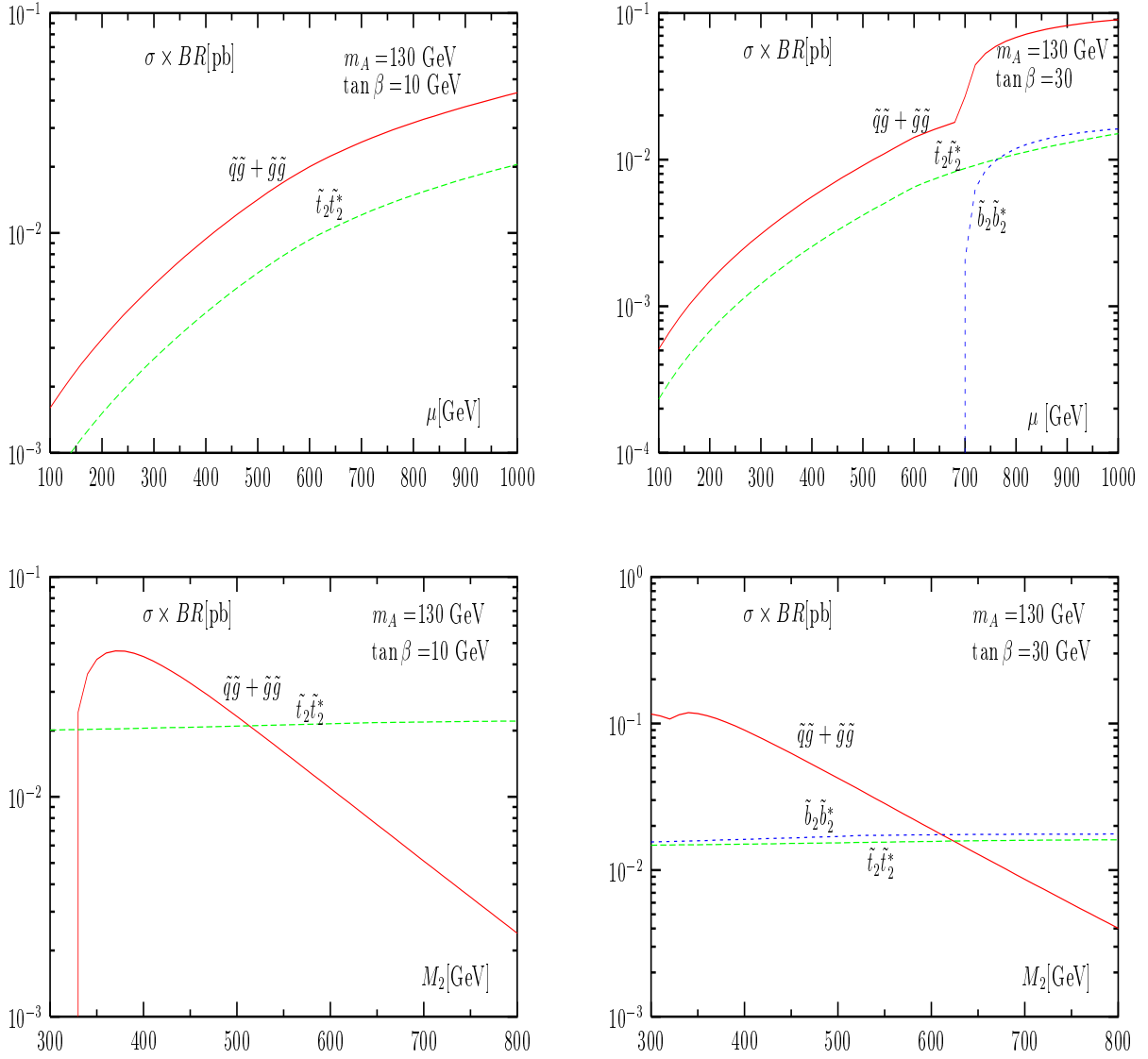,width=19cm}}
\vspace*{-6cm}
\caption{\it Cross sections times branching ratios for the production of the
pseudoscalar Higgs boson from direct decays of heavier stop and sbottom
squarks, $\tilde{t}_2 (\tilde{b}_2) \to A+ \tilde{t}_1 (\tilde{b}_1)$,  for the
values $M_A=130$ GeV and $tan\beta =10$ (left panel) and 30 (right panel).
The full lines correspond to $\tilde{t}_2, \tilde{b}_2$ squarks originating 
from decays of gluinos  produced in $pp \to \tilde{q}\tilde{g}$ and $\tilde{g}
\tilde{g}$ processes, while the dashed lines correspond to top and bottom 
squarks produced directly in the processes $pp\to \tilde{t}_2 \tilde{t}_2^*, 
\tilde{b}_2 \tilde{b}_2^*$.}
\end{center}
\end{figure}

In Fig.~6, we present the cross sections times branching ratios for the
production of {\it a single} $A$ boson in these cascade decays\footnote{We
explicitly verified that, for the CP--even neutral Higgs boson channels, the
rates are  suppressed by few orders of magnitude and are too small to be
detected at the current luminosity options at LHC.}. We have taken into account
all production channels for gluinos and squarks, including the third generation
and all possible decays of squarks, i.e. decays into $\chi$ states and quarks,
decays into gauge bosons and the possible decays into $H^\pm$ bosons.  
Representative values of the parameters are: $m_{\tilde q}=600$ GeV assuming
all soft squark masses are degenerate, $m_{\tilde g}=3 M_2$,  $A_q=1.5$ TeV and
$\tan\beta=10,30$; we fixed the pseudoscalar Higgs mass to  $M_A=130$ GeV and
varied $M_2$ or $\mu$. \s

The upper figures show $\sigma \times$BR as a function of  $\mu$  for the fixed
value $M_2=400$ GeV.  The solid lines represent the contribution from $\tilde
g$ cascade decays with the production  processes  $\tilde g \tilde g$ and
$\tilde q \tilde g$, whereas the dashed lines are for direct production of
$\tilde q_2 \tilde q_2^\ast$ pairs  [note that for $\tan\beta =10$, the sbottom
contribution is absent as the mass splitting between $\tilde b_2$ and $\tilde b_1$
is not sufficient to access  the decay $\tilde b_2 \rightarrow \tilde b_1 +
A$]. The rates increase with increasing $\mu$ values for two reasons: first,
the   decay rates $\tilde t_2 \rightarrow \tilde t_1 + A$ increase with $\mu$,
as can be seen from the $A \tilde{t}_1 \tilde{t}_2$ coupling discussed above;
secondly, a smaller phase space is accessible to stop decays into the
heavier charginos and neutralinos, at higher values of $\mu$. [Note that the
production cross sections are not sensitive to $\mu$, except for a  mild
dependence of the $\tilde t_2$ mass on this parameter.] For the higher
$\tan\beta$ case, the situation is similar, except that at a certain value of
$\mu~(\sim 700$ GeV), the splitting between $m_{\tilde b_2}$ and $m_{\tilde 
b_1}$ becomes large enough to permit the opening of the channel  $\tilde b_2
\rightarrow \tilde b_1 + A$,  which eventually contributes in both the  $\tilde
g$ cascade decay and in the direct production of $\tilde b_2$. \s

The two lower figures show $\sigma \times$BR  as functions of $M_2$ with a 
fixed value of $\mu=1000$~GeV. In the low $\tan\beta$ case, the decay of
$\tilde b_2 \to \tilde{b}_1 A$ is not accessible [as we have seen earlier],
whereas for the high $\tan\beta$ case it contributes  at the  same level as
$\tilde t_2$. Here, the rate out of the $\tilde g$ cascade decay falls very sharply
as the gluino production cross sections decrease with  increasing
$M_2=\frac{1}{3}m_{\tilde g}$. On the other hand, the stop and  sbottom
production cross sections are almost independent of $M_2$; their
slightly  rising behaviour is due to the increased decay branching ratios in 
the $A$ boson because of  the suppressed rates in  the gaugino decay channels,
which are more phase--space--disfavoured for larger $M_2$.\s

Thus, the production of $A$ bosons from direct decays of $\tilde{t},\tilde{b}$
squarks [as well as the production of $H^\pm$ bosons from the  decays
$\tilde{b} \to \tilde{t}_1 H^-$, which were discussed in Ref.~\cite{H+cascade}
and found to lead to similar results], although in general with much smaller
cross  sections times branching ratios than Higgs production from cascades
involving charginos and neutralinos, might lead  to the detection of these
particles in favourable situations. An interesting feature is that these 
processes could allow, in principle, to disentangle the $H$ and $A$ bosons in the
decoupling  limit, which is otherwise very difficult to achieve in other
processes.

\subsubsection*{2.4 Charged Higgs bosons from top decays under SUSY cascade}

The decays of the top quarks that are pair--produced at the LHC are known  to
be a very important  source of charged Higgs bosons  for $m_t > m_{H^\pm}$.
Situations permitting, as we will see in this section, cascades of SUSY
particles leading to top quarks followed by $t \to b H^{\pm}$ may also become a
significant source of $H^\pm$ bosons. \s

Charged Higgs bosons can be produced   in SUSY cascade decays via the pair
production of gluinos, top and  bottom squarks, as in eq.~(4), followed by their
cascades to top quarks, which subsequently decay to charged Higgs bosons. To
have a broad estimate of how significant this new  source of charged Higgs
particles can be, we discuss cases with the four reference scenarios Sc1--Sc4
presented previously, and calculate the rates for the production of a  {\it
single}  charged Higgs boson final state from these  cascades. To simplify our
discussion, we will not consider the possibility of  $H^\pm$ production via the
$\tilde b \tilde g$ production mechanism [which has  a very low cross section]
as well as from decays of heavier chargino and neutralino states, $\chi
\rightarrow t+X$ [which would simply increase the sample]. \s

In Fig.~7, we present the results of our study for the four scenarios with the
choice $\tan\beta=30$ and $M_{H^{\pm}} = 130$ GeV, in terms of variations of
cross sections for different production processes, times the branching ratios
for the decays $\tilde{g},\tilde{q} \to t+X$ and $t \to b H^{\pm}$. The
variations are presented with respect to $\mu$ for Sc1 and Sc2 and with respect
to $M_2$ for Sc3 and Sc4.  We do not present the results for a lower  $\tb$ value
since it can be estimated in a straightforward manner  from the one we are
analysing: while  the production cross sections and decay  branching ratios of
SUSY particles remain almost constant, the final $\sigma \times$BR drops by a
factor  $\sim 1/\tan^2\beta$ [i.e. by a factor $\sim 9$ when going from
$\tan\beta=30$ to 10]  because of the suppressed Yukawa coupling, $g_{H^\pm tb}
\propto m_b \tb$ [the component of this coupling that is proportional to $m_t$
is suppressed by  a factor  $1/\tb$]. Therefore, the total $\sigma \times$BR in
the $\tb=10$ case is one order of  magnitude smaller than in Fig.~7, while the
relative magnitude of the  various contributions is approximately the same.
Note that, in each figure,  the long arrow along the $y$ axis represents the
$H^\pm$ production cross section  via direct $t \bar t$ production. \s

The $\sigma \times$BRs  are suppressed by roughly  an order of magnitude when
going from Sc1 to Sc2 owing to an increased gluino mass and consequently larger
squark masses. The variations with respect to $\mu$ present in these scenarios
are mainly determined by the resulting variation of the $\tilde{t} t  \chi_i^0$
and $\tilde{b} t \chi_i^\pm$ couplings. For small $\mu$ values [$\ll M_2$] the
lighter neutralinos  and chargino have masses $m_{\chi_{1,2}^0},
m_{\chi_1^\pm} \simeq \mu$ while they are  higgsino--like, yielding
larger branching ratios in those $\tilde g$  decay channels and also for
$\tilde{t}_1 \to t \chi_j^0$($j=$1--4).  However, at a certain value of $\mu$, 
$\chi_3^0, \chi_4^0$ become so heavy that they are not accessible in
stop decay; this results in a smaller or vanishing BR($\tilde t_1 \rightarrow 
t \chi^0_{3,4}$) value. A further increment of $\sigma \times$BR beyond
this $\mu$ value in Sc1 is due to a gradual increment of the gaugino content 
of $\chi_{1,2}^0$. \s

\begin{figure}[htbp]
\begin{center}
\vspace*{-1.2cm}
\centerline{\epsfig{file=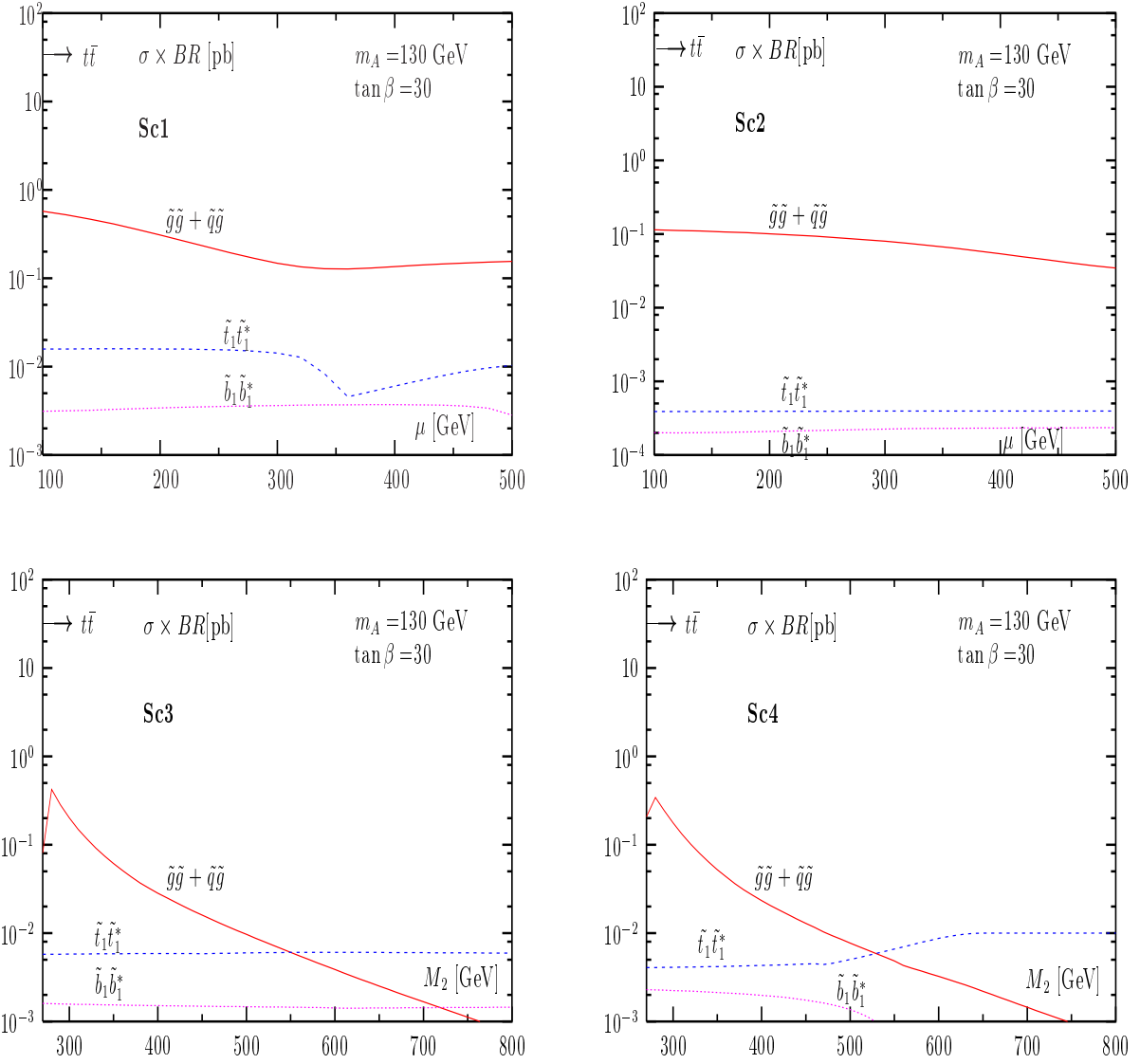,width=19cm}}
\vspace*{-6cm}
\caption{\it Cross sections times branching ratios for charged Higgs boson
production from decays of top quarks originating from cascade decays of
strongly interacting SUSY particles in the scenarios Sc1 to Sc4 and for the values $M_A=130$
GeV and $tan \beta=30$. For $tan \beta=10$, the trend is similar except that
the rates are one order of magnitude smaller. The arrows on top of the figures
correspond to the  cross section for $gg/q\bar{q} \to t\bar{t}$ times the
branching ratio for the decays $t \to H^+b$ and $\bar{t} \to H^- \bar{b}$.} 
\end{center}
\end{figure}

The profile of the $\tilde{g}$ contribution in Sc1 roughly follows that of
$\tilde{t}_1 \tilde{t}_1^*$, because the same decay mode $\tilde{t}_1 \to t
\chi_1^0$ is involved there, to end up with top quarks.  The flatness of
$\sigma \times$BR  for  $\tilde{b}_1 \tilde{b}_1^*$ production then indicates
the low sensitivity  of the $\tilde{b} \chi^{\pm} t$ coupling to $\mu$, which
is due to the small $m_b$ value.  For Sc2 the trend is similar, except for the
fact that now, the  kinks do not appear in the frame. This is due to an 
increased value  of $m_{\tilde{g}}$ [and hence, of $m_{\tilde q}=1.2 m_{\tilde
g}$]  which effectively  pushes $m_{\tilde t_1}$ to higher values, enabling
access to all neutralino states in its decay $\tilde t_1 \rightarrow t
\chi_j^0$ for the entire range of $\mu$. \s

In Sc3 and Sc4 the contributions from $\tilde{g} \tilde{g}$ and $\tilde{g} 
\tilde{q}$ production drop naturally with  increasing $M_2=\frac{1}{3}M_{\tilde
g}$, as a result of the smaller production cross section for heavier gluinos.
The peaks at  about $M_2 \simeq 270$ GeV indicate the onset of the two--body
decay of a gluino,  $\tilde g \rightarrow t \tilde t_1$. While going from Sc3
to Sc4, we see that the situation does not change much, except for a slight
increase of $\sigma \times$BR  in the $\tilde{t}_1$ pair--production  case and
a drop in the $\tilde{b}_1$ pair--production case beyond 500 GeV in Sc4. The
flatness of the $\tilde{t}_1 \tilde{t}_1^*$ contributions in Sc3 indicates that
the main contribution is coming from  $\tilde{t}_1 \to t \chi_i^0$ where $i=$1
and 2, i.e. from the lighter neutralinos with masses close to $\mu$. In the
case of sbottoms, the main decay mode seems also to be $\tilde{b}_1 \to t
\chi_1^-$, this chargino being a higgsino. \s

With $\mu=1$ TeV for Sc4, the decay $\tilde{b}_1 \to t \chi_1^-$ gradually drops
with increasing $M_2$, leading to a decrease of available phase--space
[here, the lighter chargino is gaugino--like] and gets closed completely for a
high $M_2$ value. However, in the stop case there is an increase in  $\sigma
\times$BR at about  500 GeV for these contributions, which indicates the onset
of a gradual increment of the branching ratio $\tilde t_1 \rightarrow t
\chi_1^0$, at the expense of other channels including decays into higher
neutralinos and charginos as they are getting closer to phase--space suppression
with the increase of $M_2$ for a value of $\mu$ as high as 1 TeV. \s

Thus, for the large value of $\tb=30$ considered  and for Sc1  and
Sc2, this additional contribution to single charged Higgs  production 
may even become comparable to the one from the cascades involving charginos and
neutralinos, in a rather conservative estimate [the cross sections times
branching ratios being at the level of 0.1 pb]. For Sc3 and Sc4, these new
contributions can only  be significant for low gluino masses and naturally drop
sharply with the increase in the latter. As mentioned  already, and contrary to
the processes discussed in section 2.2, the event  rates decrease quadratically
with $\tb$ and, for instance, are an order of magnitude smaller for $\tb=10$.
\s

In short, for not too heavy a gluino, $m_{\tilde{g}} \lsim 1$ TeV, and large 
values of $\tan\beta$, $\tb \gsim 20$, top quarks coming from SUSY particle 
cascade decays may contribute heavily to charged Higgs boson final states in
some regions of the MSSM  parameter space. This would always strengthen the
signal, although one has to be rather meticulous about the backgrounds [in
particular, it would not be an easy task to separate these events from the ones
where the $H^\pm$ bosons originate from the top quarks,  which are directly
produced]. \s

\subsection*{3. Detector simulation analysis} 

In the previous section, the production mechanism of MSSM Higgs bosons through 
strongly interacting SUSY particle cascade decays has been extensively
discussed for several SUSY scenarios. The important feature of this mechanism is
the quasi--independence of the production cross sections times branching ratios
on  the value of $\tan \beta$. This is in contrast to the strong $\tan \beta$ 
dependence of the direct production mechanisms for the heavier charged and
neutral MSSM Higgs bosons, which in the standard processes [$gg \to H/A$ or
$gg/q\bar{q} \to b\bar{b}H/A$ for the neutral Higgs bosons and $gg \to
t\bar{b}H^-$ or $gb \to H^-t$ for the charged Higgs particles] lead to a
significant number of events only for larger value of this parameter, for which
the $Hb\bar{b}, Ab\bar{b}$ and $H^\pm tb$ couplings are sufficiently enhanced.
On the other hand, the masses of the MSSM Higgs bosons  accessible in the SUSY
cascade decays will probably be limited to the range $M_A \lsim 250$ GeV by
phase--space considerations. Therefore, one might  hope to cover the lacuna in
the discovery reach for the heavy MSSM Higgs bosons at the LHC, which is left
after including the SM [and possibly sparticle] decay modes of these particles,
i.e. for low pseudoscalar Higgs boson masses, 120 GeV $\lsim M_A \lsim $ 200 
GeV and moderately small $\tan \beta$ values, $3 \lsim \tb \lsim 10$.\s

In this section, we will present a first analysis that will demonstrate the
detection potential of the MSSM Higgs bosons in SUSY particle cascade decays.
As already mentioned, the subject is rather involved and in the
present paper, we will simply perform a fast Monte--Carlo simulation of the
production signal and the various SM  and SUSY backgrounds for $H,A$ and
$H^\pm$ boson production via squark and gluino decays through the big and small
cascades. [We will not discuss direct decays of top and bottom squarks into
Higgs particles or the production of $H^\pm$ bosons through top quark
decays].   For illustration, we will focus on the case of $\tb =5$ and
$M_A=150$ GeV, which leads to a light Higgs boson with a mass of $M_h \simeq 
110$ GeV [since $h$ is not yet SM--like for this set of parameters, this mass
value is not yet excluded by LEP2 searches] and a charged Higgs boson with a 
mass  $M_{H^\pm} \simeq 170$ GeV [a value for which the decays $t \to H^+b$ 
would be strongly suppressed], in the scenarios discussed in section 2. The
cross sections times branching ratios for these scenarios have been displayed in
Fig.~5.  \s

The signal and the background events are generated with {\tt HERWIG 6.4}
\cite{HERWIG}. The SUSY spectrum is calculated\footnote{Some of the sparticle
and Higgs boson masses as well as some of the branching fractions for the 
cascade decays obtained with this program are slightly different from those
obtained in our numerical discussion of the previous section. However, this
discrepancy does not alter the main features of our analysis and can be 
resolved [i.e. one can obtain a similar spectrum] by slightly changing the
inputs.} with {\tt ISASUSY} 7.58 \cite{ISASUSY} and then interfaced to  {\tt
HERWIG} using the {\tt ISAWIG} package \cite{ISAWIG}. To be as realistic as
possible, we also simulate some aspects of the CMS detector using {\tt CMSJET}
4.801 \cite{CMSJET}, which contains a parametrized description of the CMS
detector response.  The effects of event pile--up at the LHC have not been
included, but are expected to be small for our analysis.  The features that
will allow the distinction between the Higgs boson signals from the SM and SUSY
backgrounds will be discussed in the next two subsections, and we will argue
that the SM backgrounds can be efficiently suppressed.  However, a more
difficult task will be to distinguish the signals from other SUSY cascade decay
processes since all squark and gluino production at the LHC will end up in
lightest neutralinos and fermions through gaugino and slepton decays.  Hence, a
good understanding of the nature of these SUSY backgrounds will be needed prior
to any search for MSSM Higgs bosons in cascade decays.

\subsubsection*{3.1 Neutral Higgs bosons}

We start by describing a first--analysis strategy that will allow the
observation of the neutral $h,H$  and $A$ bosons produced in the
squark/gluino cascades. Two decay modes of  these neutral Higgs particles are
promising, namely $A,H \rightarrow b \bar{b}$, with a branching ratio around
90\%, and $A,H \rightarrow \tau \tau$, with a  branching ratio around 10\% [the
branching ratios are slightly smaller in the  case of the lighter $h$ boson].
Because of the dominant branching rate, we will now focus on the $b\bar{b}$
decays, although the $\tau^+ \tau^-$  decay may have some advantages due to the
lower multiplicity of  $\tau$--jets with respect to $b$--jets in the cascade
background.  Eventually both modes should be studied since the ratio of their
signals  could be used to support the evidence for a Higgs signal in the SUSY
cascade decays.\s

In all scenarios, the $h,H$ and $A$  bosons originate from the $\tilde{g}
\tilde{g}$, $\tilde{q}\tilde{q}$, $\tilde{q}\tilde{q}^*$ or $\tilde{q}
\tilde{g}$ production processes through the chain illustrated in eq.~(1) or
(2). In  Sc1 and Sc3, only the first chain is allowed, i.e. only the
``big cascade" mechanisms are at work. In Sc2, the light $h$ boson
can be produced through the ``little cascade" mechanism of eq.~(2). In
Sc4, also the $H,A$ bosons can be produced in the little cascades. In all
cases the background consists of the same $\tilde{g}\tilde{g}$,
$\tilde{q}\tilde{q}$, $\tilde{q}\tilde{q}^*, \tilde{q}\tilde{g}$ processes,
where the squarks and gluinos decay into neutralinos and charginos that do not
lead to Higgs bosons in their subsequent decay chains. \s

We will study the features of the signal and the backgrounds in order to
determine a set of  selection criteria that should serve the following generic
purposes:

\begin{itemize}
\vspace*{-3mm}

\item They should make sure that the events can pass the CMS on-line trigger 
levels with a high efficiency.
\vspace*{-2mm}

\item They should reduce the SM backgrounds to the level where they are 
negligible with respect to the SUSY backgrounds.
\vspace*{-3mm}

\end{itemize}

These two purposes are closely related, since the triggers are designed to 
reject SM jet backgrounds. In contrast to SM processes, cascade decay Higgs 
events contain a large number of very energetic jets and a large transverse 
missing energy. By selecting only events with these features using the 
criteria that we will propose,  trigger efficiencies of more than 90\% can be
reached \cite{DAQTDR}. Additional cuts intending to
suppress the SUSY background {\it i.e.} the SUSY events not containing any Higgs
bosons, will depend on the details of the kinematics
of this background. In the present paper, to be as unbiased as possible,   we
will not adapt any special cuts to reject this SUSY background. After
suppression of the SM background, we will merely look for resonances in the bb
invariant mass spectrum for the four scenarios, comparing  this spectrum to its
equivalent in the case where Higgs boson production in the SUSY cascades is not
allowed.\s

We will now discuss the kinematical distributions of the SUSY cascades and the SM
background in order to obtain a recipe that allows us to suppress the latter
while preserving the former. With 'SUSY cascades' we mean the production of all 
sparticles according to cross sections obtained from the four scenarios as 
described in the previous section.
\s

The concrete parameter choices for these scenarios are [the gluino mass 
$m_{\tilde{g}}$ has been slightly changed in Sc3 and Sc4  to obtain a spectrum 
that is closer to the one discussed in section 2, while the common slepton
mass  has been fixed in all four scenarios to the value $m_{\tilde{l}}$ = 500
GeV]:

\begin{itemize}
\vspace*{-2mm}

\item Sc1:  $M_2=2M_1= 200$ GeV, $\mu = 300$ GeV, $m_{\tilde{g}} = 600$ GeV and 
$m_{\tilde{q}} = 720$ GeV. 
\vspace*{-2mm}

\item Sc2: $M_2= 2M_1= 300$ GeV, $\mu = 450$ GeV, $m_{\tilde{g}} = 900$ GeV and 
$m_{\tilde{q}} = 1080$ GeV.
\vspace*{-2mm}

\item Sc3: $M_2= 2M_1= 350$ GeV, $\mu = 150$ GeV, $m_{\tilde{g}} =
1200$ GeV and $m_{\tilde{q}} = 800$ GeV. 
\vspace*{-2mm}

\item Sc4: $M_2= 2M_1= 350$ GeV, $\mu = 1000$ GeV, $m_{\tilde{g}} = 
1200$ GeV and $m_{\tilde{q}} =800$ GeV. 
\vspace*{-2mm}
\end{itemize}

As SM backgrounds, we have only simulated the main $t\bar{t}$ process. Also
other QCD processes [$b\bar{b}$, $c\bar{c}$ and light quark production]  will
contribute to the SM background; however, it is very difficult to produce a 
reliable QCD sample in this case, since extreme kinematical fluctuations are
necessary for this type of background to be within the signal cuts. Therefore
we have not included the QCD background, being confident that  any set of
criteria that is able to suppress the $t\bar{t}$ background will also be
effective against the other QCD background.  We recall that {\tt ISASUSY} 7.58
and {\tt HERWIG} 6.4 are used for the event generation and that the detector
aspects were simulated using {\tt CMSJET} 4.801 \cite{CMSJET}, which contains
fast parametrizations of the CMS  detector response and a parametrized track
reconstruction performance based on {\tt GEANT} simulations for $b$--quark tagging
\cite{CMSIM}. Results for signal and background are summarized in
Figs.~8--11. \s

\begin{figure}
\begin{center}
\epsfig{file=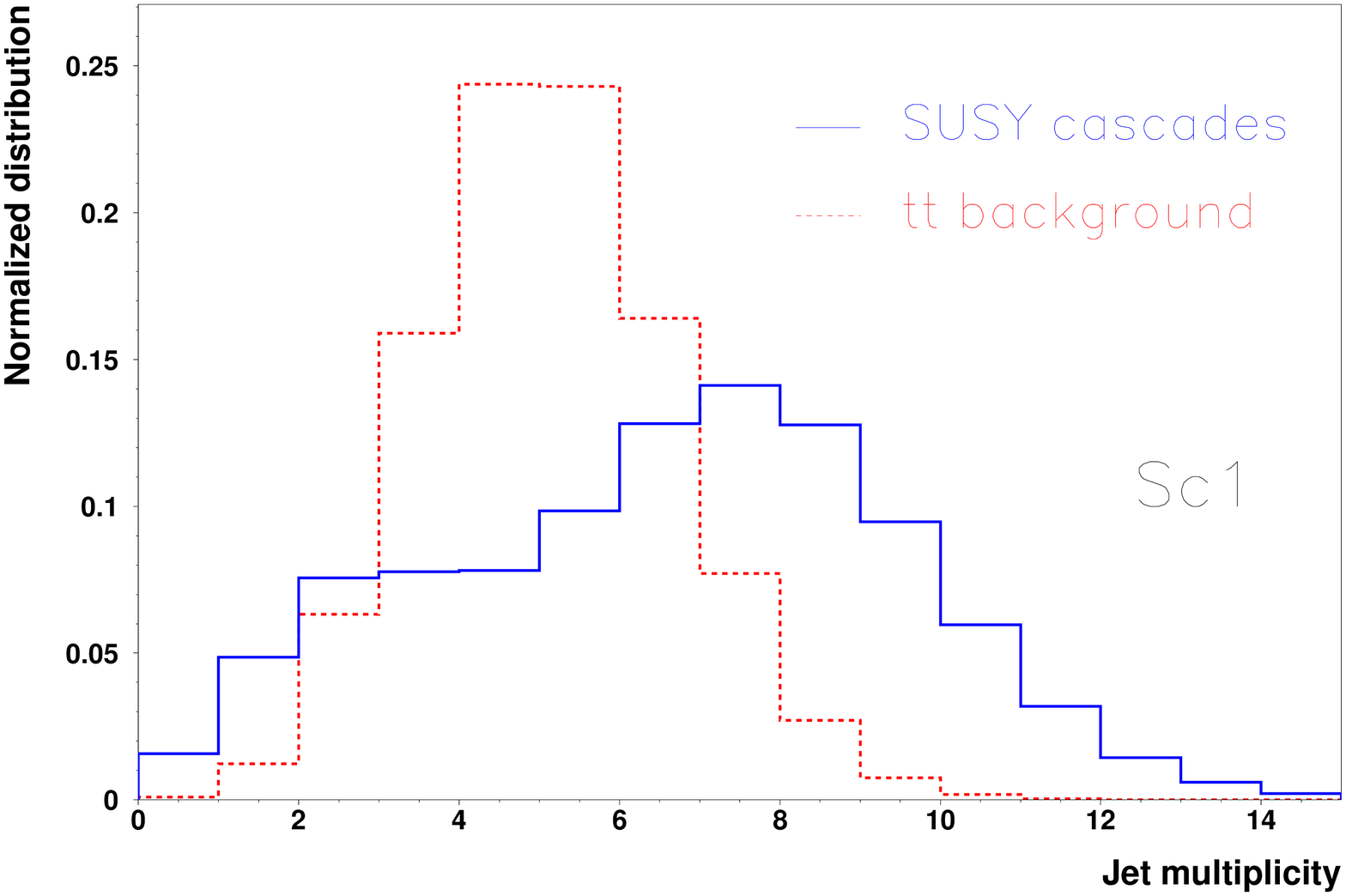,height=85mm,width=75mm}
\epsfig{file=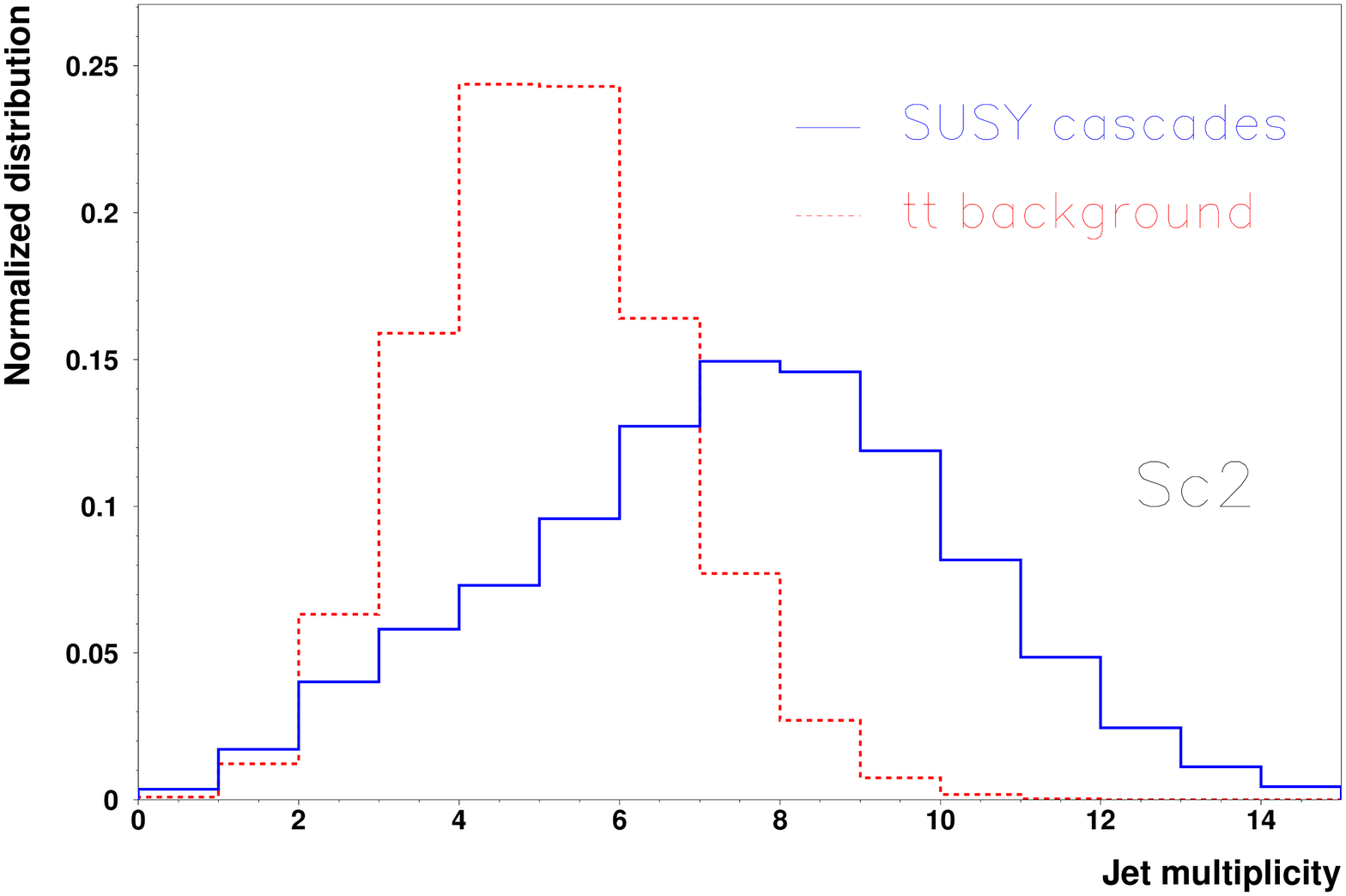,height=85mm,width=75mm}
\end{center}
\begin{center}
\epsfig{file=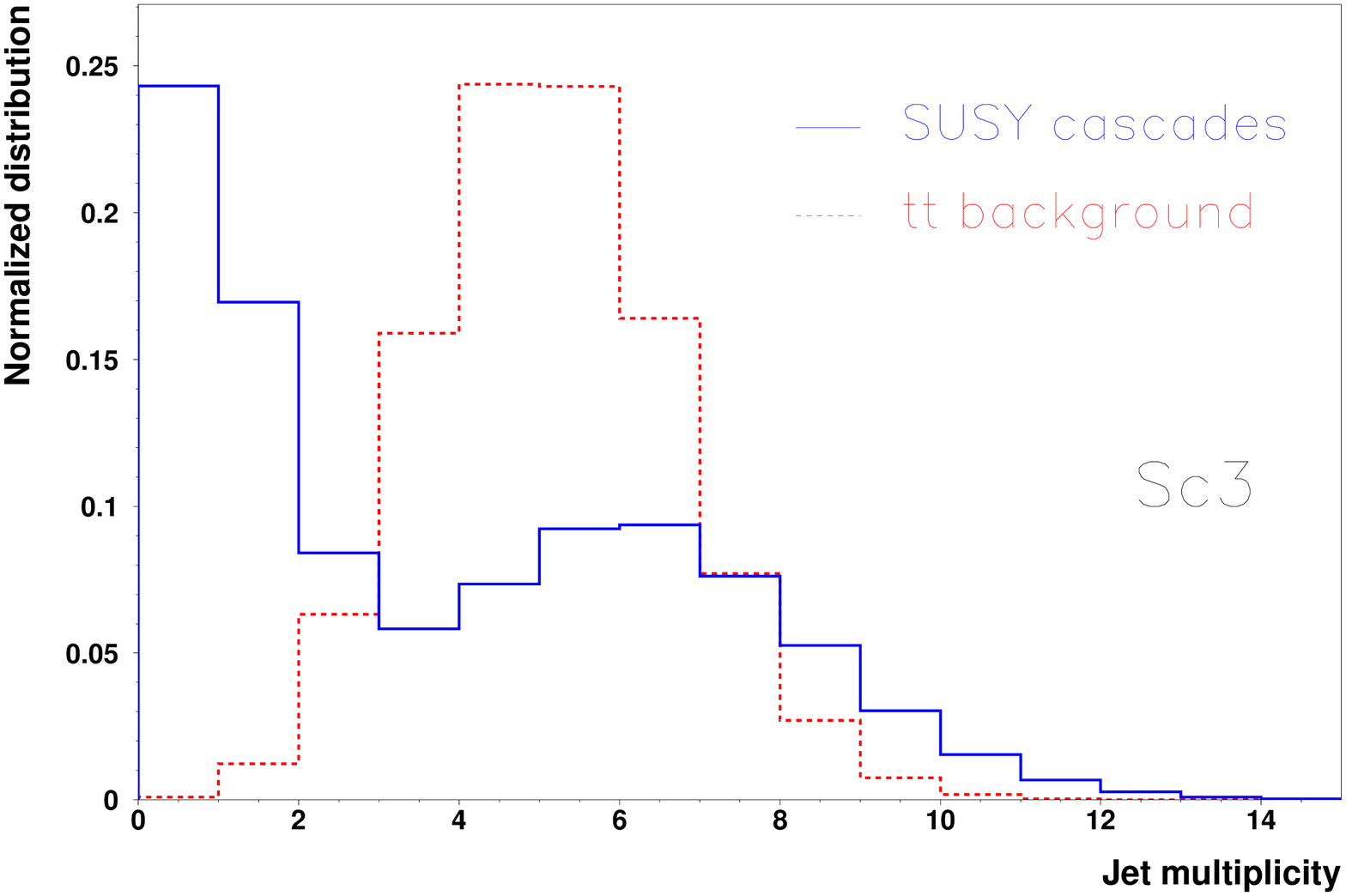,height=85mm,width=75mm}
\epsfig{file=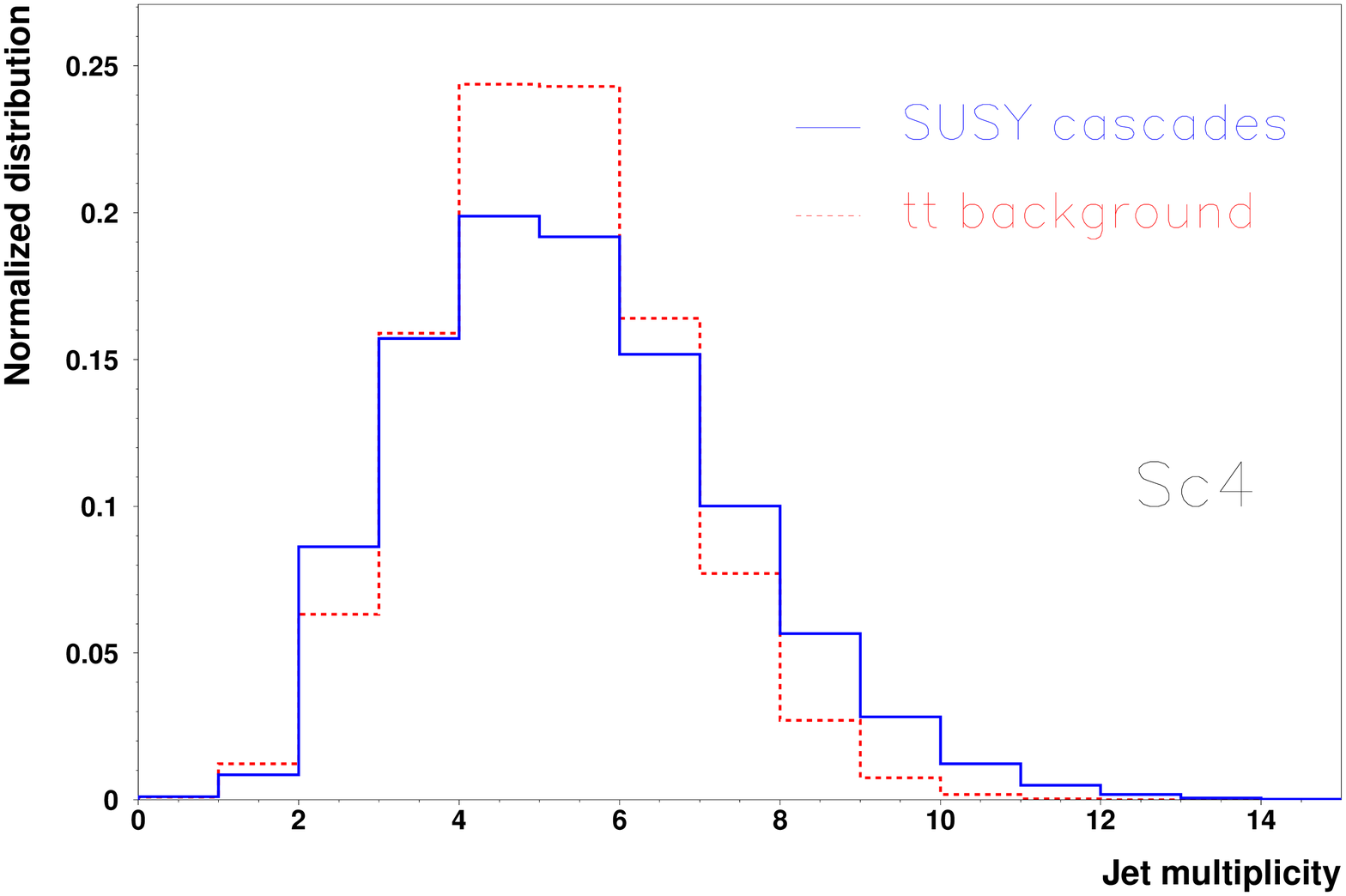,height=85mm,width=75mm}
\end{center}
\caption{Normalized distribution of the jet multiplicity in the SUSY cascades (full line)
and the SM $t\bar{t}$ background (dashed line) events, for the four scenarios.
Typically, the SUSY cascades events contain more jets than the SM background events.}
\label{fig:distr1}
\end{figure}
\begin{figure}
\begin{center}
\epsfig{file=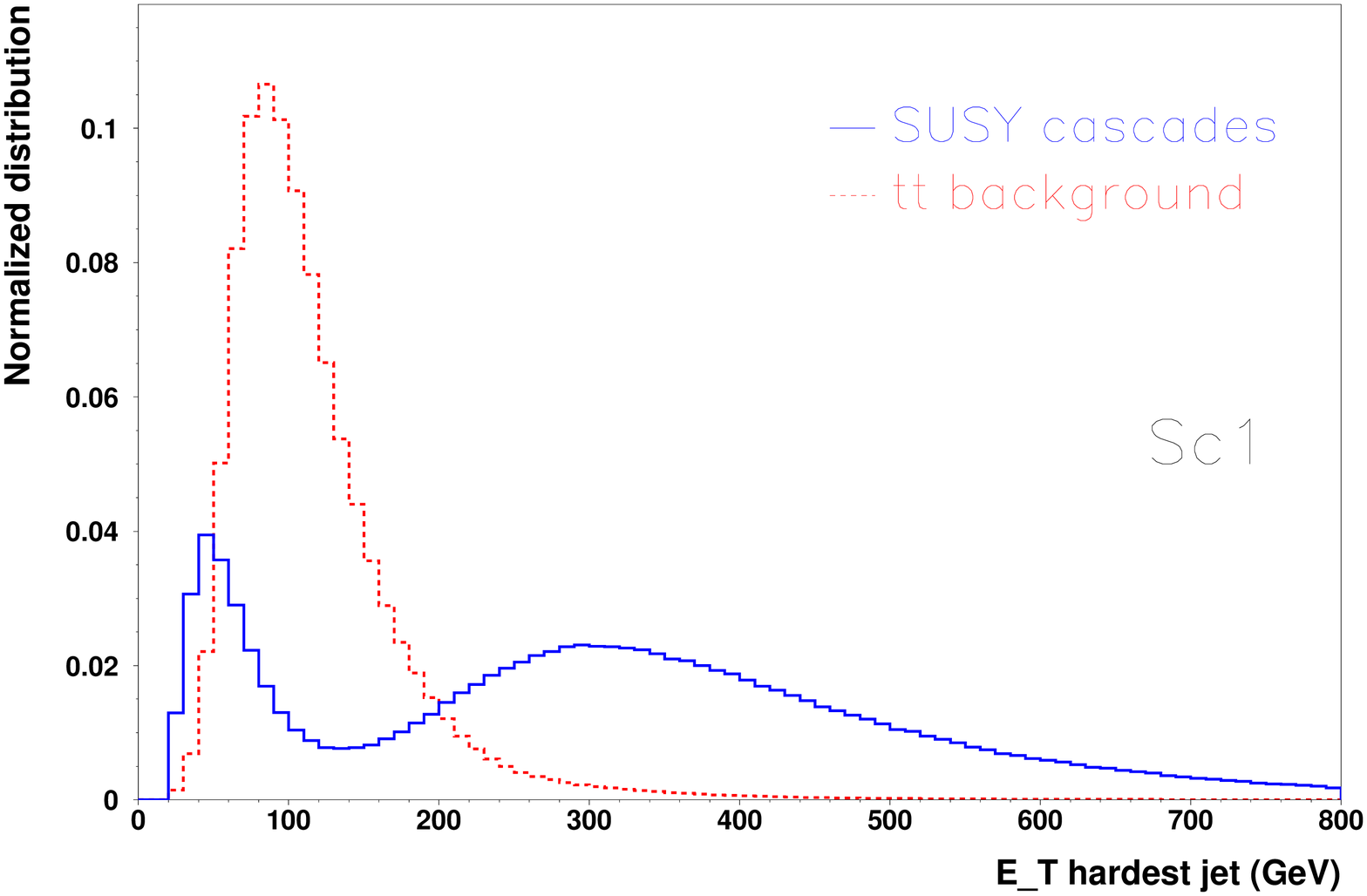,height=85mm,width=75mm}
\epsfig{file=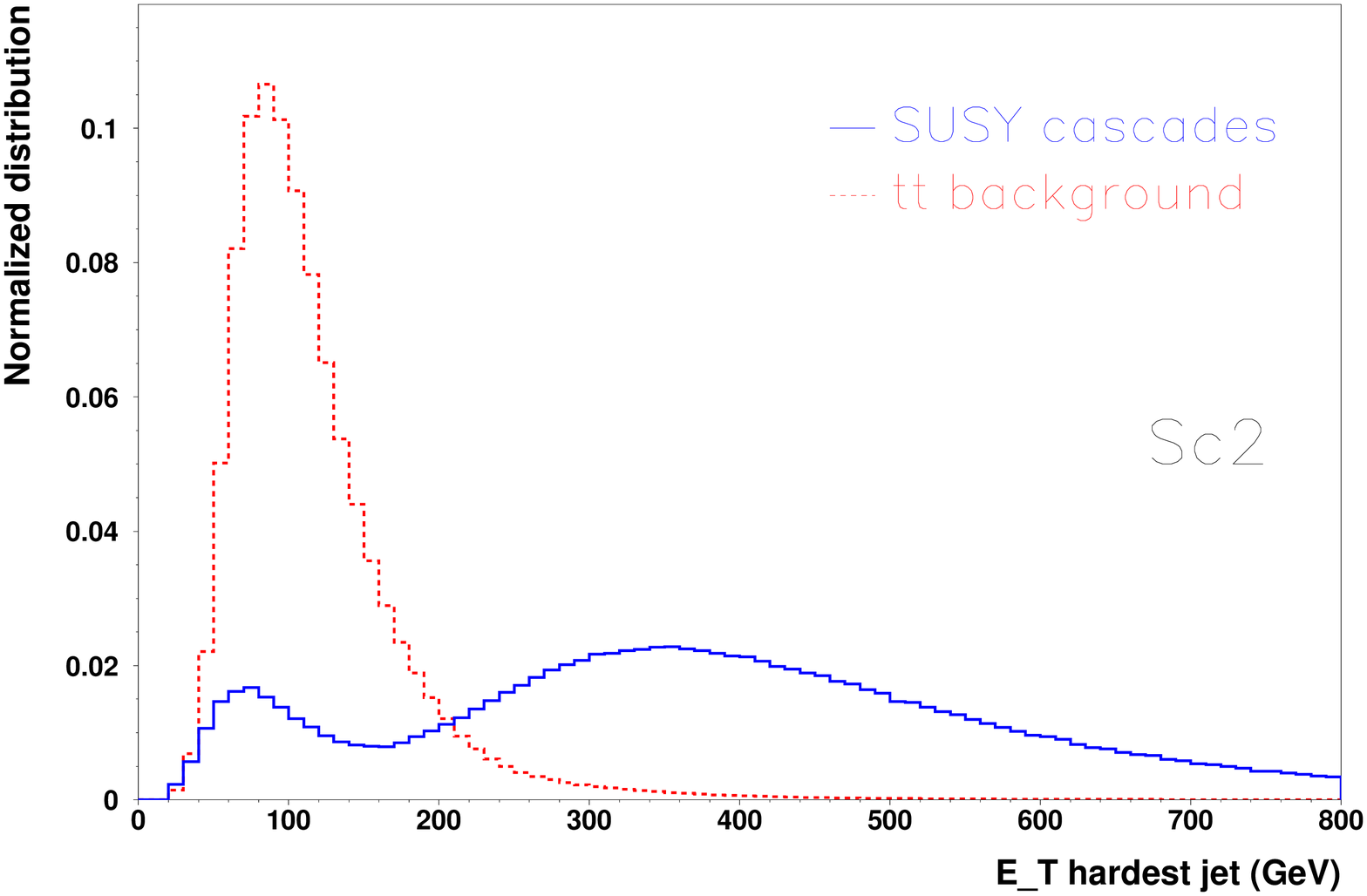,height=85mm,width=75mm}
\end{center}
\begin{center}
\epsfig{file=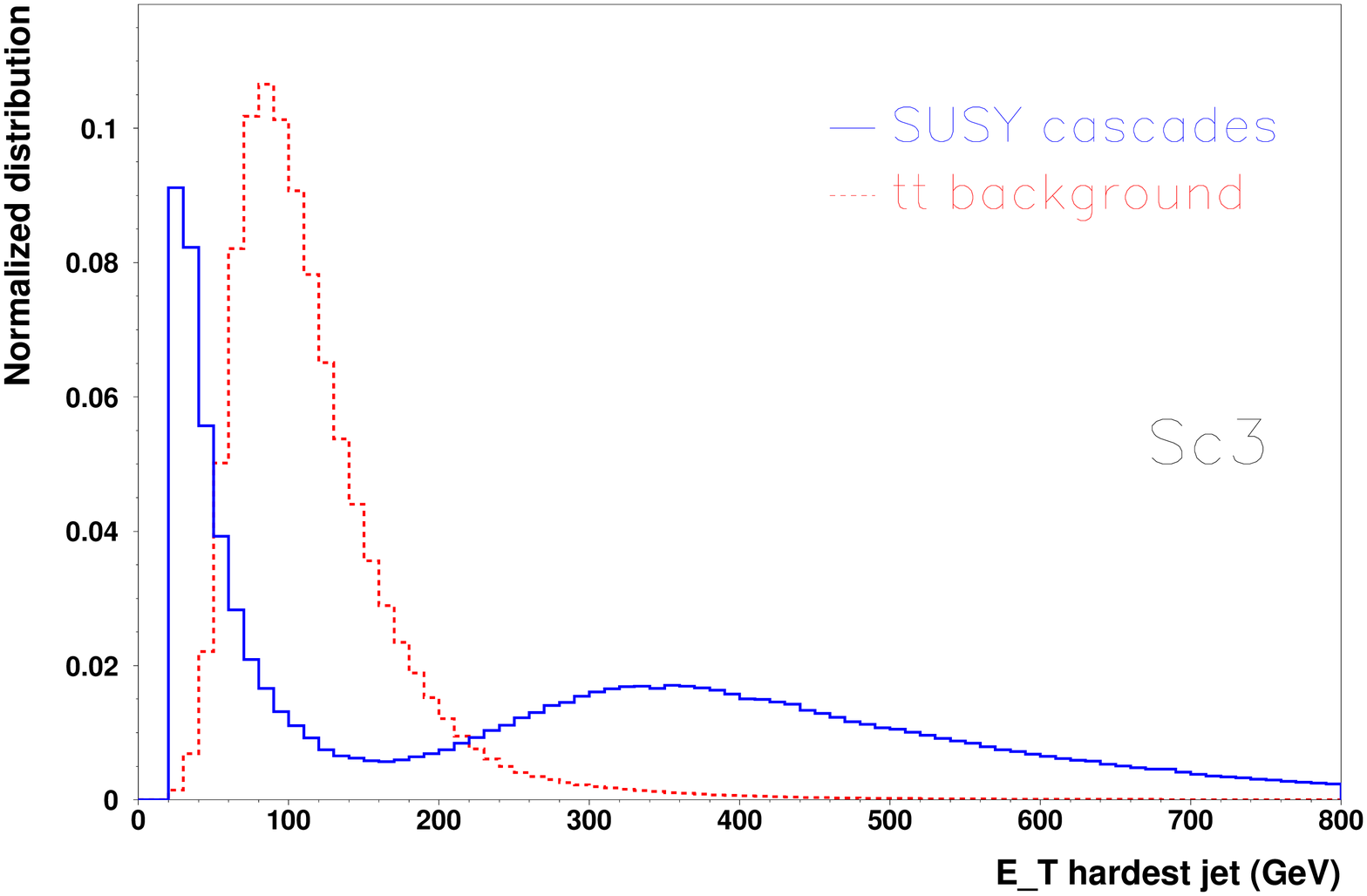,height=85mm,width=75mm}
\epsfig{file=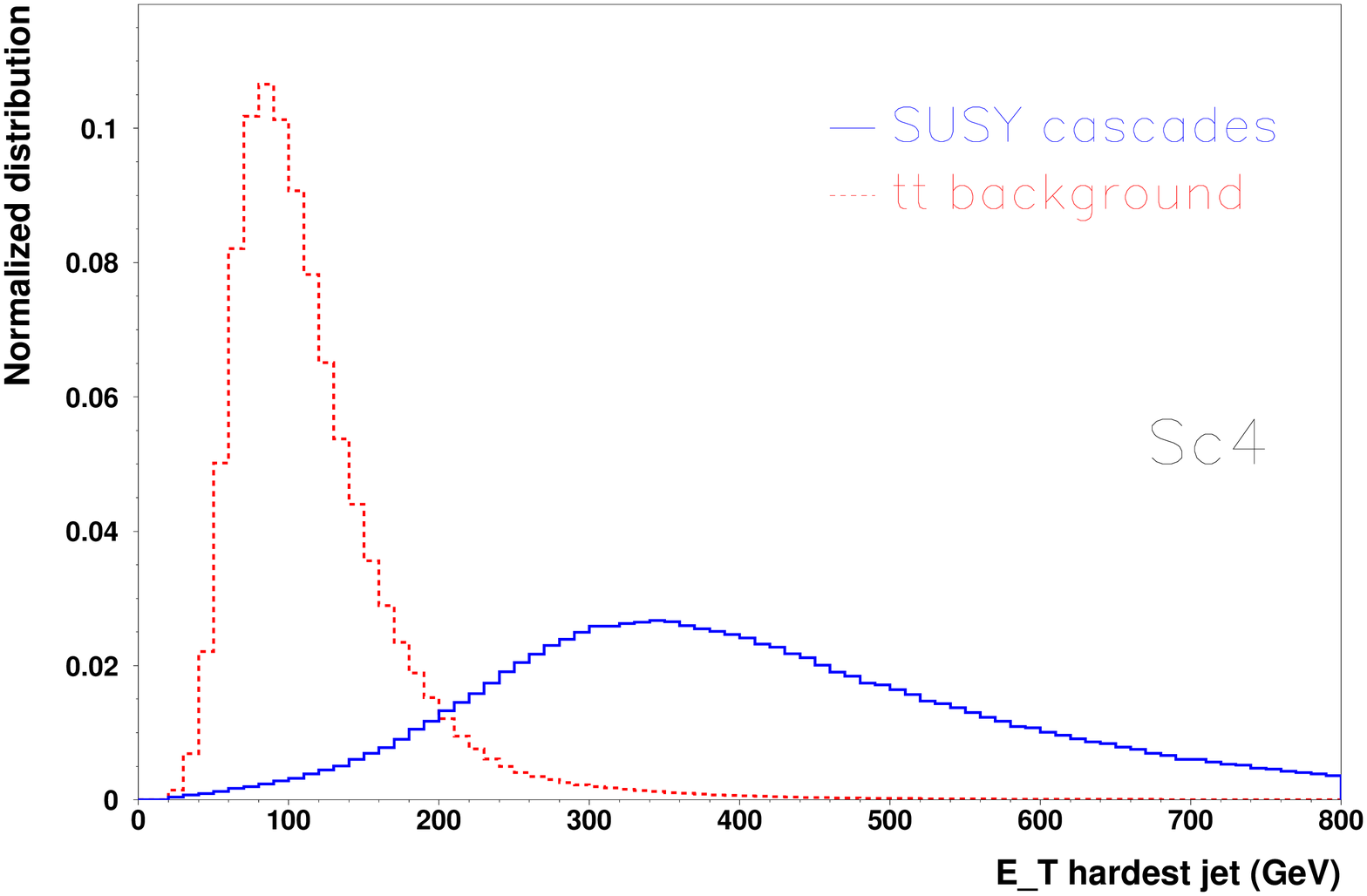,height=85mm,width=75mm}
\end{center}
\caption{Normalized distribution of the $E_T$ of the hardest jet in the SUSY
cascades (full line)
and the SM $t\bar{t}$ background (dashed line) events, for the four scenarios. 
The distributions peak around 100 GeV for the SM background
and above 300 GeV for the SUSY cascades.}
\label{fig:distr2}
\end{figure}
\begin{figure}
\begin{center}
\epsfig{file=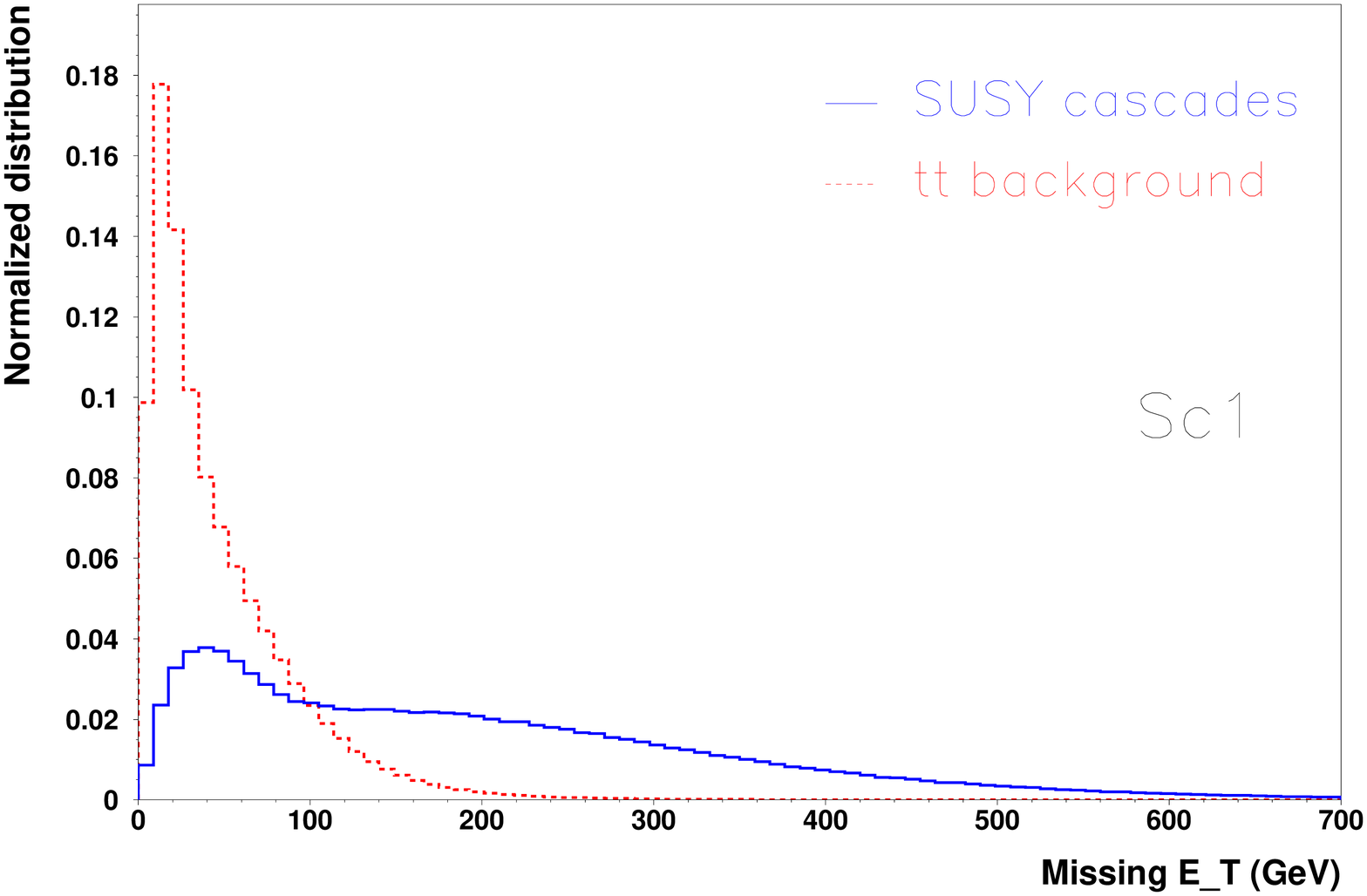,height=85mm,width=75mm}
\epsfig{file=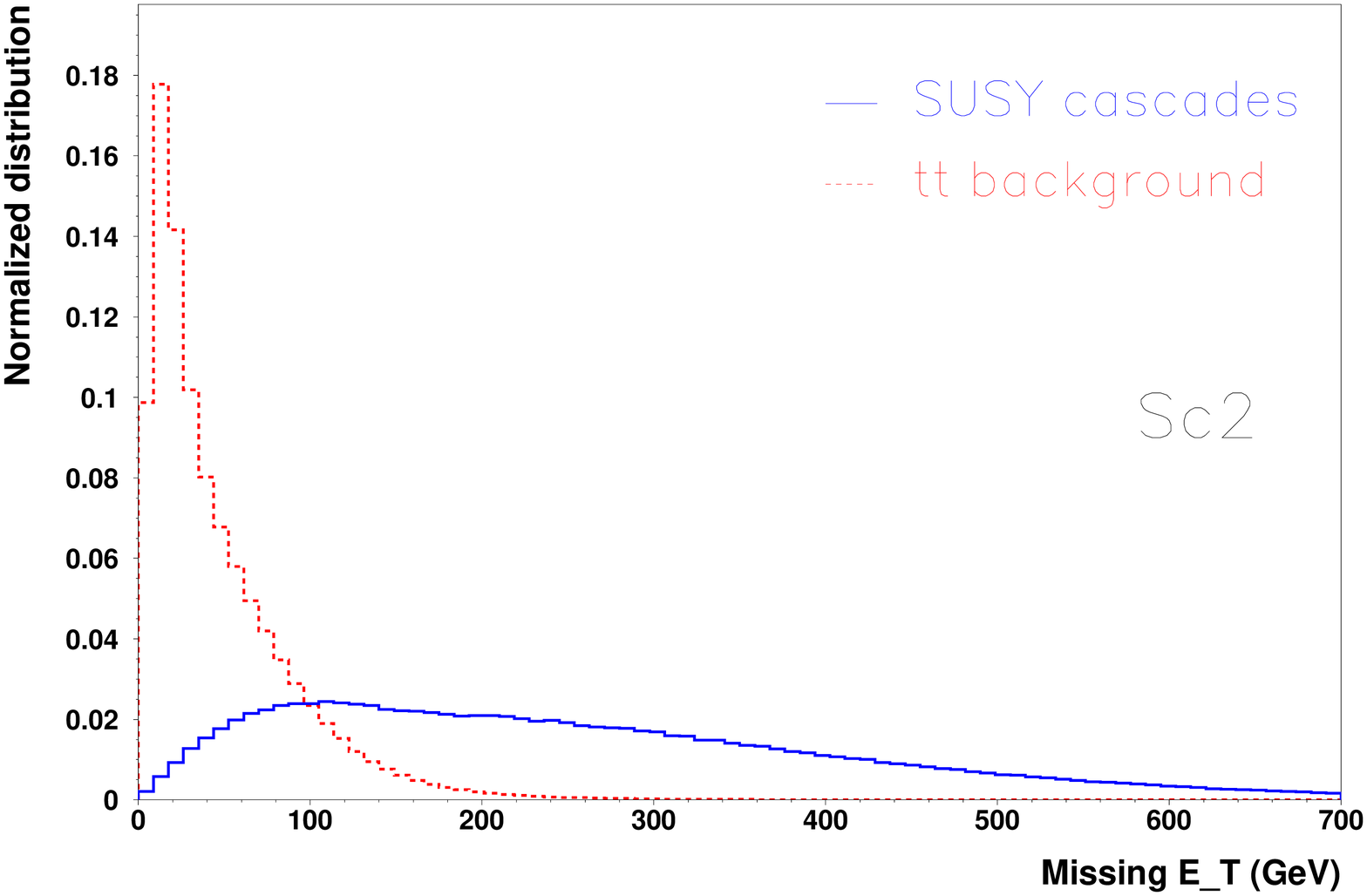,height=85mm,width=75mm}
\end{center}
\begin{center}
\epsfig{file=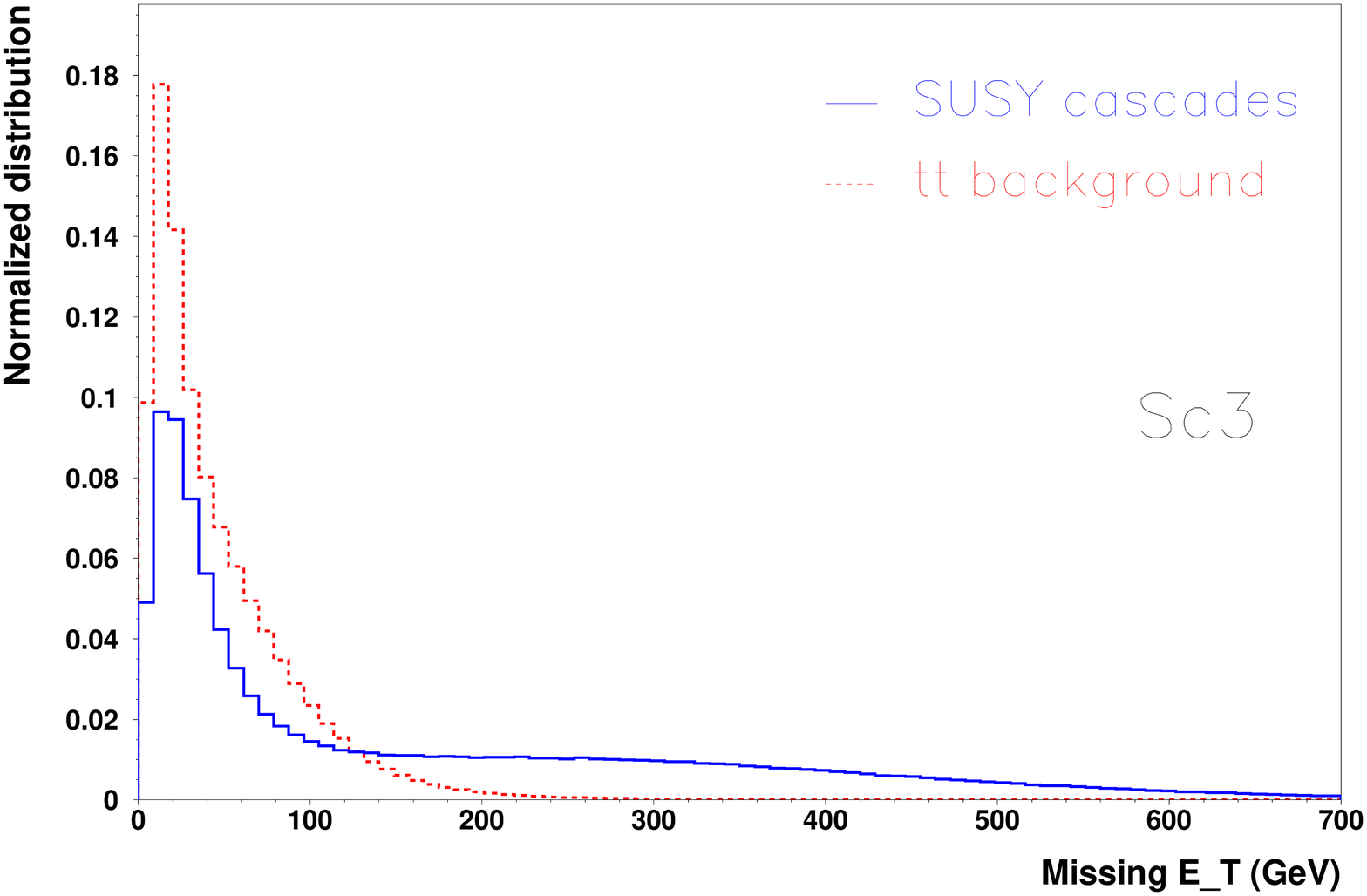,height=85mm,width=75mm}
\epsfig{file=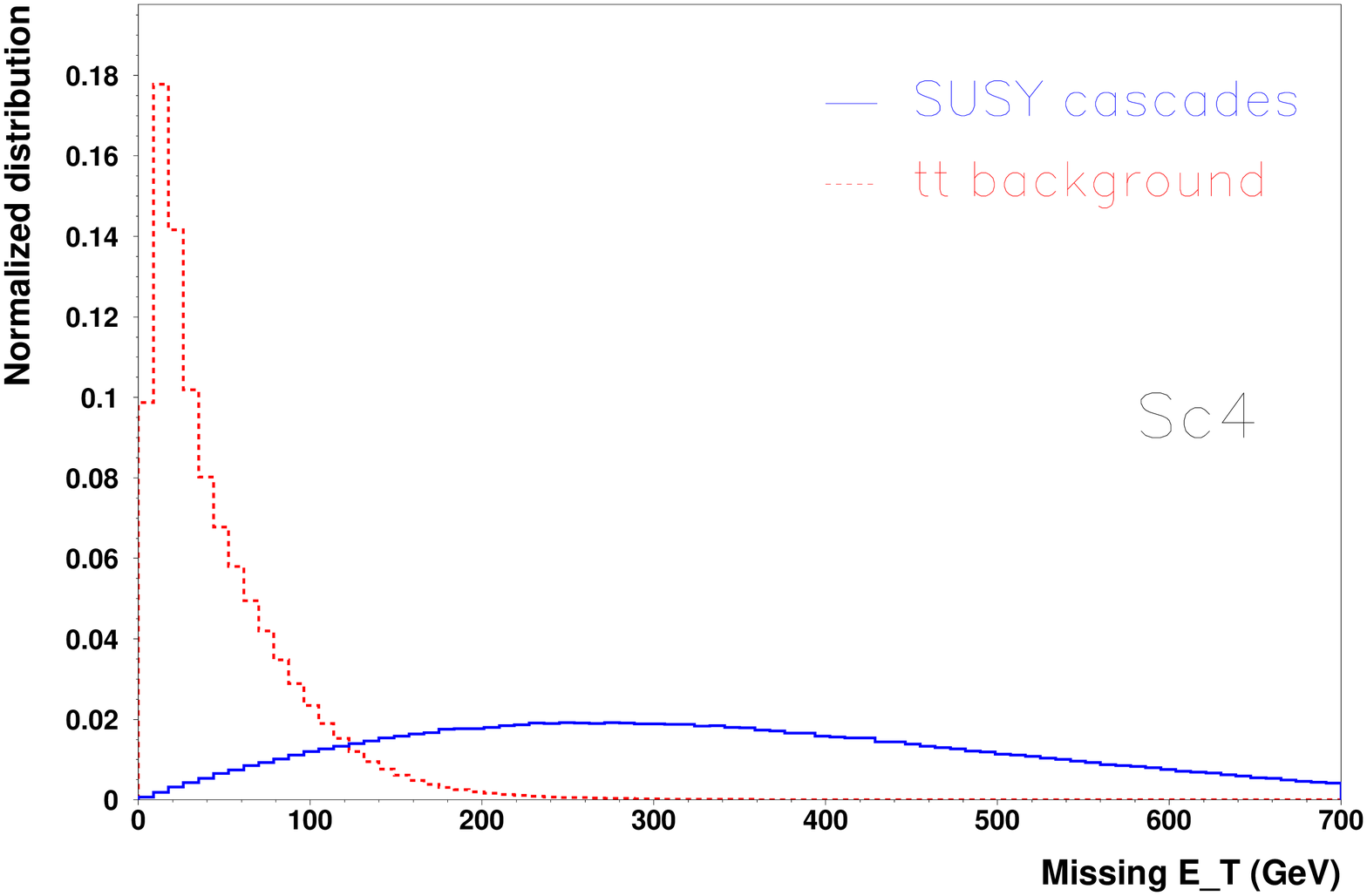,height=85mm,width=75mm}
\end{center}
\caption{Normalized distribution of the transverse missing energy in the SUSY
cascades (full line)
and the SM $t\bar{t}$ background (dashed line) events, for the four scenarios. 
The distribution of the SM background peaks at low values and drops steeply,
while the SUSY cascades has a very broad distribution reaching very high values.}
\label{fig:distr3}
\end{figure}
\begin{figure}
\begin{center}
\epsfig{file=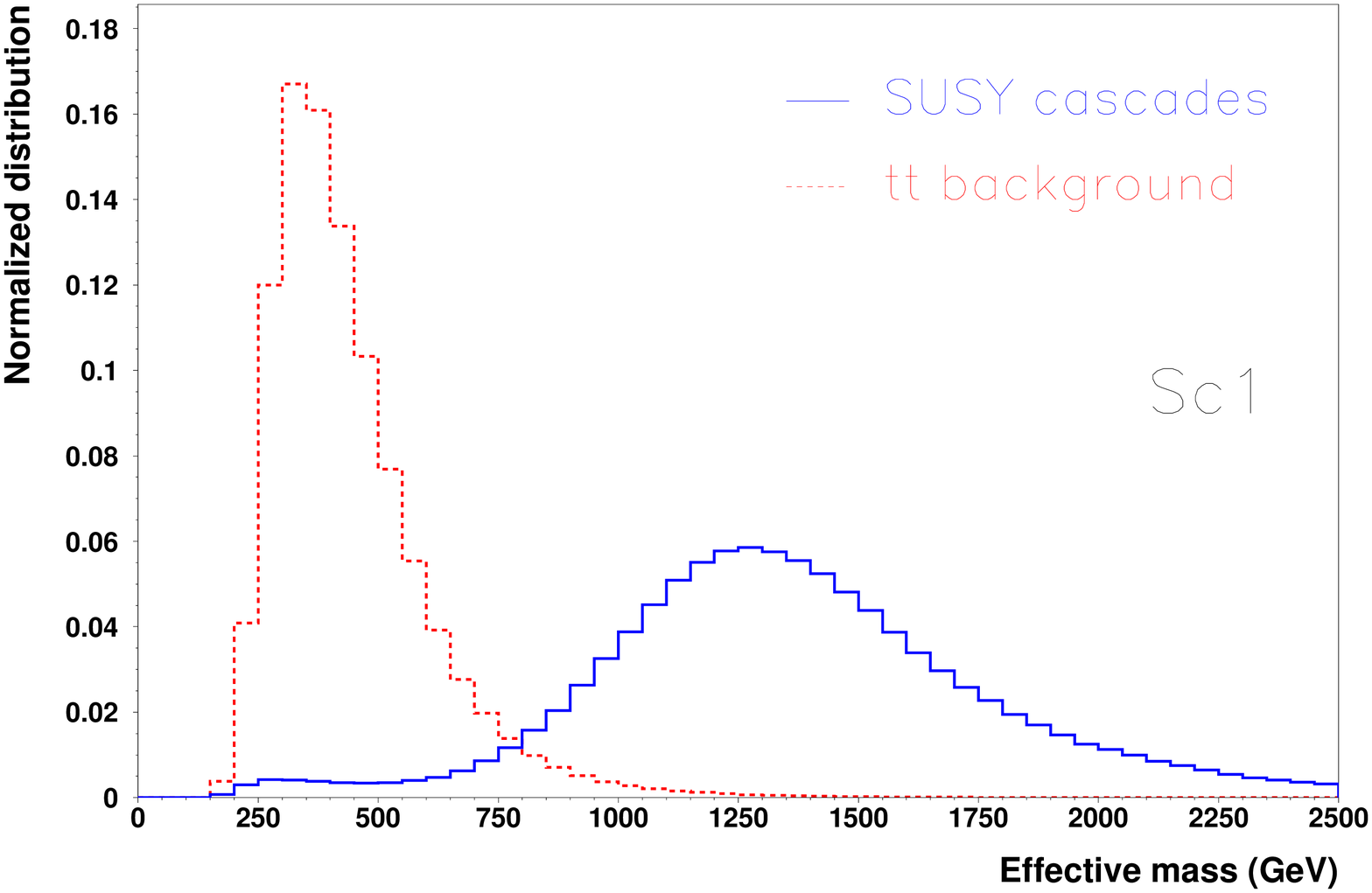,height=85mm,width=75mm}
\epsfig{file=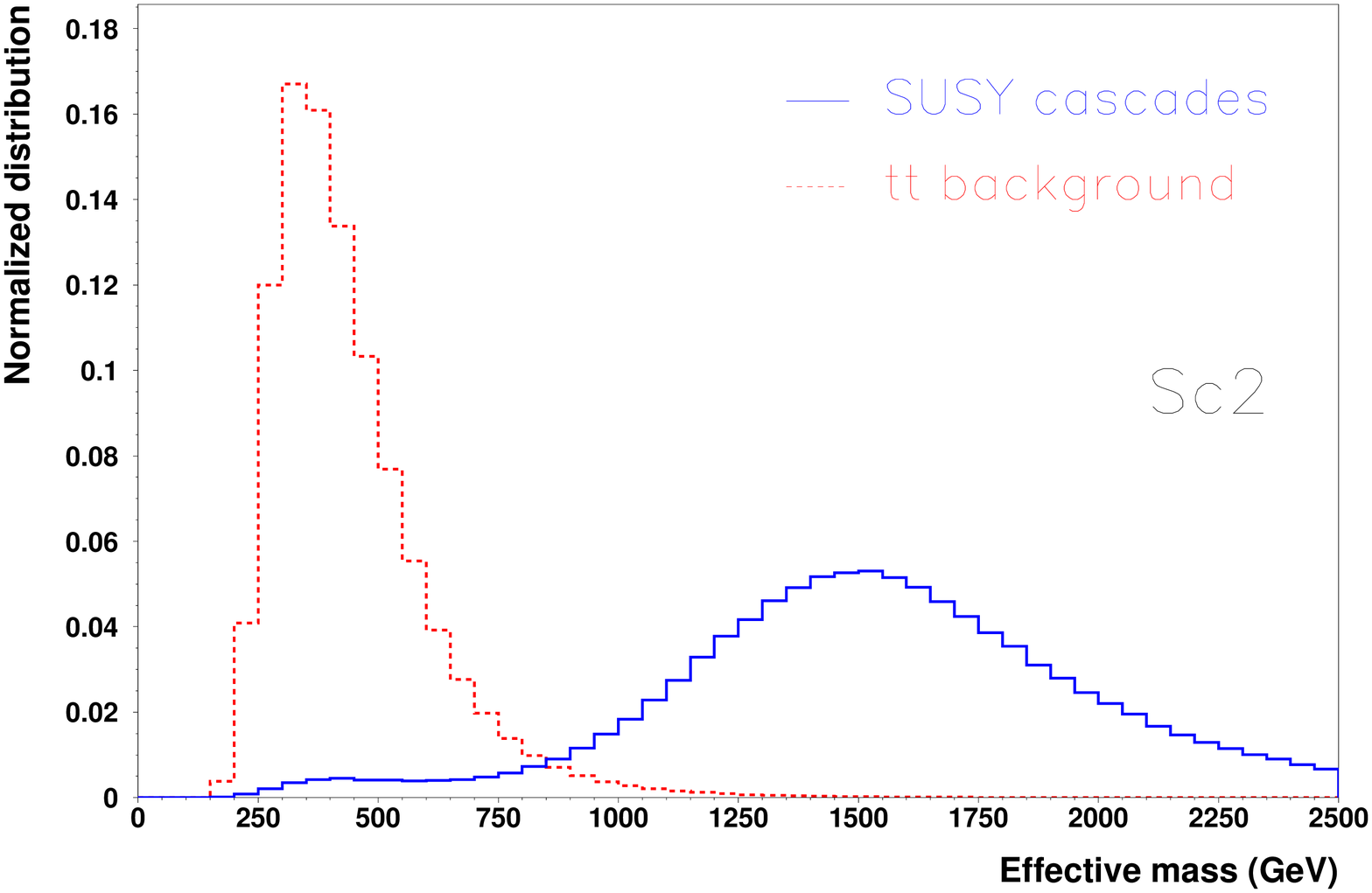,height=85mm,width=75mm}
\end{center}
\begin{center}
\epsfig{file=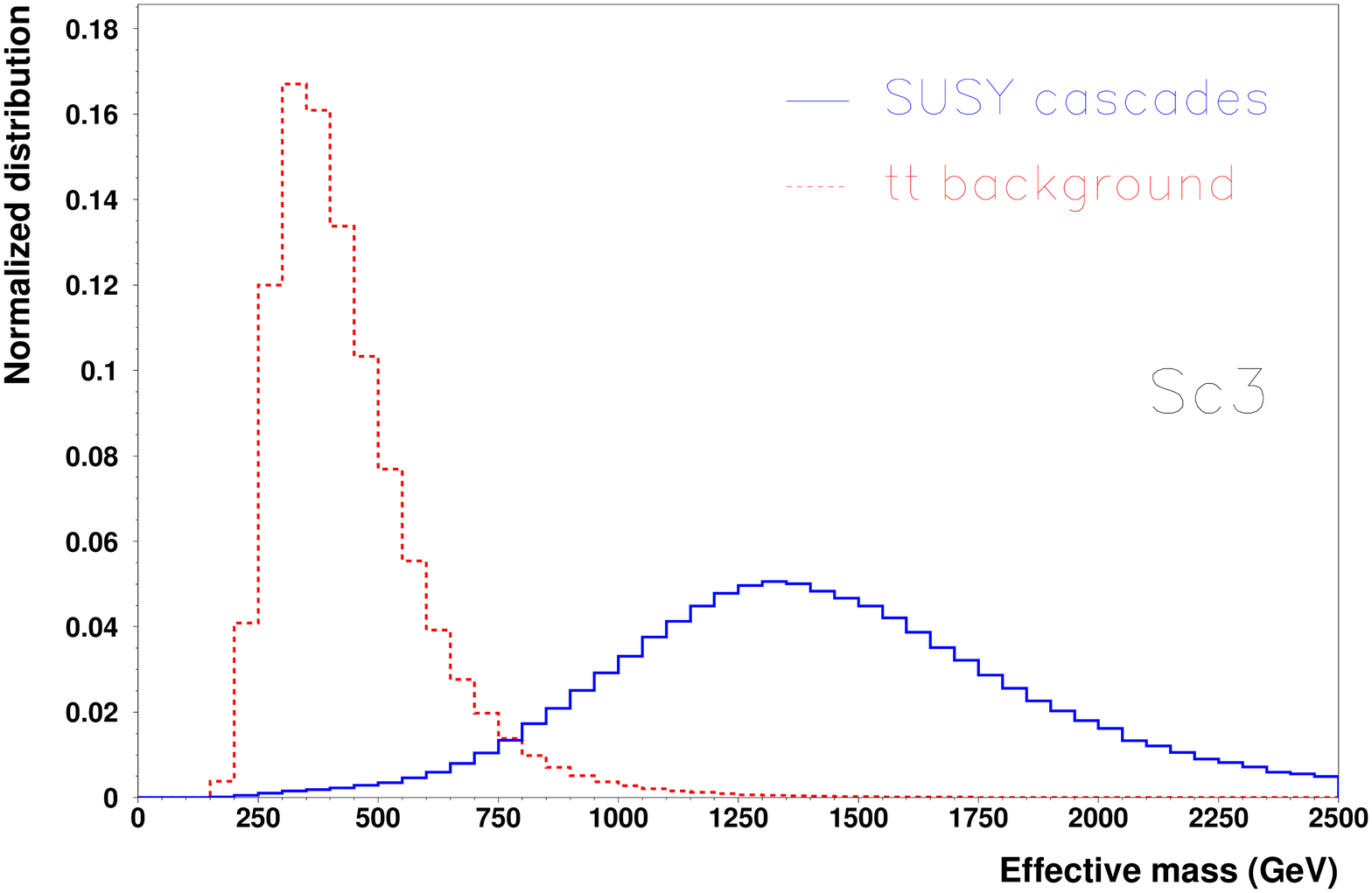,height=85mm,width=75mm}
\epsfig{file=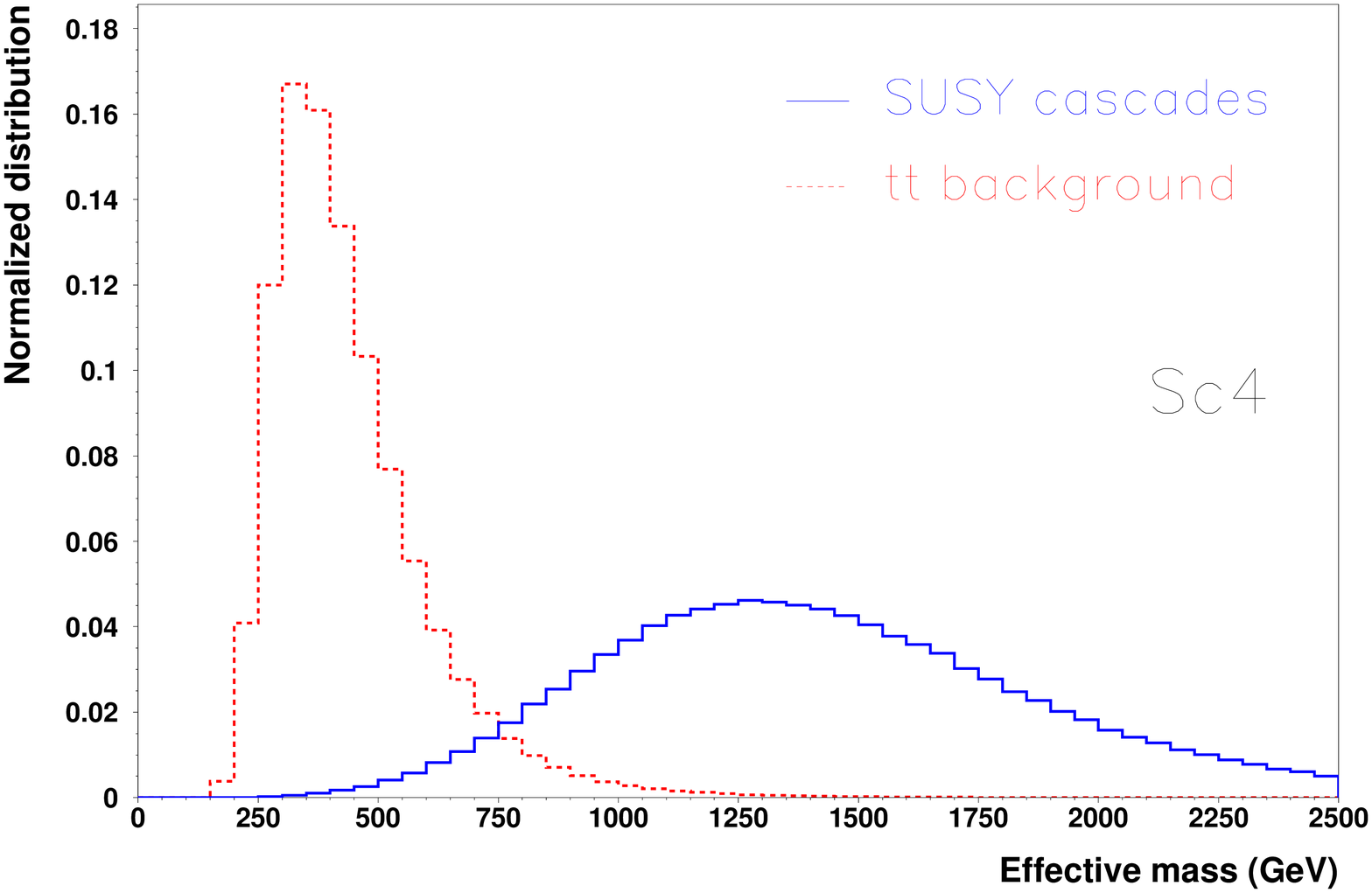,height=85mm,width=75mm}
\end{center}
\caption{Normalized distribution of the effective mass consisting of the sum of the $E_T$ of the jets
and the $E_T^{\rm miss}$ in the SUSY cascades (full line)
and the SM $t\bar{t}$ background (dashed line) events, for the four scenarios. 
The distributions of the signal and the SM background are clearly separated in 
all scenarios.}
\label{fig:distr4}
\end{figure}

In Fig.~\ref{fig:distr1} the jet multiplicity is shown, for the SUSY cascades
and the SM $t\bar{t}$ background, for each of the four scenarios. The
first two scenarios have a jet multiplicity distribution that peaks around 8
jets, while the two last scenarios show somewhat lower jet multiplicities. This
is linked to the hierarchy of squark, gluino and gaugino masses  as described
in section 2. For the SM $t\bar{t}$ background, the distribution peaks around 5
jets.  Requiring a large number of jets in the event will enhance the sparticle
production versus the SM background in most scenarios. \s

In Fig.~\ref{fig:distr2}, we can see the transverse energy of the hardest jet
in the event, for the SUSY cascades and the SM $t\bar{t}$ background
events, in each of the four scenarios. The distribution peaks around 100 GeV for
the SM background and above 300 GeV for the SUSY cascades. Demanding the
$E_T$ of the hardest jet to be above $\sim$ 300 GeV will therefore strongly
suppress the  SM background. \s

Fig.~\ref{fig:distr3} shows the transverse missing energy in the events,  for
the SUSY cascades and the SM $t\bar{t}$ background, in each of the four
scenarios. The distribution of the SM background peaks at low values and drops
steeply, while the SUSY cascades have a very broad distribution reaching
very high values. Again, requiring a large $E_T^{\rm miss}$ in the events will
strongly favor the sparticle production with respect to the SM background. \s

In Fig.~\ref{fig:distr4}, the variable $E_T^{\rm total}$ is shown, consisting of
the sum of the $E_T$ of all the jets and the $E_T^{\rm miss}$ in the event. 
The distributions of the SM $t\bar{t}$ background and the SUSY cascades
are clearly separated in each of the four scenarios.  Selecting only events in
which $E_T^{\rm total}$ is above $\sim$ 1200 GeV  will help suppress the
SM background. This selection cut in itself is very effective, 
however it becomes somewhat redundant if applied after the two previous cuts
on $E_T^{\rm jet}$ and $E_T^{\rm miss}$. \s

Combining the above observations allow us to strongly suppress the Standard
Model   background while preserving most of the SUSY cascade events. As a next
step,  we will study the $b\bar{b}$ invariant mass spectrum and look for
resonances of the $h$ (110 GeV), $A$ (150 GeV) and $H$ (160 GeV) Higgs bosons.
In order to perform this, we will demand that there be at least two  jets in
the event carrying a significant $b$--tag and having $E_T$ values  compatible
with the ones originating from a Higgs boson. As mentioned before, no other cut
will be made to suppress the SUSY  background in view of the uncertain nature
of this background. 
We can thus summarize the selection strategy as follows:

\begin{itemize}
\vspace*{-3mm}

\item[$i)$] We require the event to contain at least 5 jets.
\vspace*{-3mm}

\item[$ii)$] The hardest jet in the event should have an energy $E_T$ 
larger than 300 GeV.
\vspace*{-3mm}

\item[$iii)$]  The transverse missing energy $E_T^{\rm miss}$ should be larger than 
150 GeV.
\vspace*{-3mm}

\item[$iv)$]  The variable $E_T^{total}= \sum E_T^{\rm jets}+E_T^{\rm miss}$ should 
have a value larger than 1200 GeV.
\vspace*{-3mm}

\item[$v)$] The event should contain at least two $b$--jets with 45 $<$ $E_T$
$<$ 120 GeV. In order for the jets to be $b$--tagged, we demand that both jets
contain at least two tracks with a significance of the transverse impact
parameter $\sigma(ip)=ip_{xy}/\Delta ip_{xy}>$ 3.
\vspace*{-3mm}
\end{itemize}

If more than two jets in the event fulfill the last condition, the $b\bar{b}$
invariant mass is calculated using the two $b$--jets that are closest to each
other in $\eta$--$\phi$ space. Applying the above condition $\sigma(ip)>$ 3
leads to a b--tagging efficiency of about 40\%, with a mistagging probability
of less than 0.5\%. \s

For each of the four scenarios, we will apply this selection strategy to two 
different squark/gluino samples: one where all decays are allowed (SUSY signal)
and another one where the chargino/neutralino decays into Higgs bosons are
switched off (SUSY background). By overlapping the $b\bar{b}$ invariant mass
distribution  obtained with both samples, we can study the effects of Higgs
bosons in the SUSY cascades in an unbiased way.  The main results are 
summarized in Figs.~\ref{fig:final1}--\ref{fig:final4}, assuming a
luminosity   of $\int {\cal L}=30$ fb$^{-1}$.\s

Fig.~\ref{fig:final1} shows the $b\bar{b}$ invariant mass spectrum for the SUSY
signal overlapped with the SUSY cascade background for Sc1. Both signal
and background feature a $Z$ boson peak. In our fast simulation code, the
``measured" $M_{b\bar{b}}$ is typically underestimated by about 10\% compared to
the ``true" $M_{b\bar{b}}$. However, the $Z$ peak seems to be shifted to lower 
values by more that. 
This is due the large presence of $\chi^0_2 \rightarrow b\bar{b} \chi^0_1$ decays 
in the cascades, the $b\bar{b}$ invariant mass distribution of which features a kinematical
edge around 80 GeV. This edge interferes with the $Z$ peak and leads to the
apparent peak around 70 GeV.
There are two more, relatively small, peaks visible  in the
signal $M_{b\bar{b}}$ spectrum: one around 100 GeV and one around 140 GeV. The
first one corresponds to the presence of the $h$ boson while the second one
corresponds to the $A,H$ bosons. In this scenario, all Higgs bosons are being
produced in the big cascades, i.e. they originate from heavy
chargino/neutralino decays. In order to obtain a more convincing S/B ratio,
this scenario requires additional selection cuts to suppress the SUSY background.
In this figure, also the Standard Model $t\bar{t}$ background 
is shown. Clearly it is small compared to the SUSY background, 
justifying our motivation to neglect the QCD background.
\s

Fig.~\ref{fig:final2} shows the $M_{b\bar{b}}$ distributions for Sc2 and
one can see that a large signal peak is coming from $h \rightarrow b \bar{b}$ 
decays.  This is due to the fact that the $h$ boson is being produced in the
little cascades  [$\chi^0_2 \rightarrow h \chi^0_1$ has a branching fraction of
96\%],  while the $H$ and $A$ are only produced in the big cascades. Therefore
the $H$, $A$ peak is less visible.  Moreover, there is a broad smearing due to
the large combinatorial background coming from the wrong pairing of $b$--jets
since  the $H$ or $A$ state is often accompanied by a $\chi^0_2$, which decays
$\chi^0_2 \rightarrow h \chi^0_1$, leading to four $b$ quarks in the event. In 
the background, only the $Z$ boson peak can be seen.\s

In Fig.~\ref{fig:final3}, the result is shown for the third scenario.  There,
as can be seen in Fig.~3, the  production rates for all   Higgs bosons are
rather high. There are no little cascades, and  the  rates for $h$ and $H$, $A$
production are therefore not very different. A double Gaussian fit clearly
shows the corresponding peaks, while in the background only the  $Z$ boson peak
is visible. \s

In Fig.~\ref{fig:final4}, the $b\bar{b}$ invariant mass spectrum is shown for
Sc4. In this case, not only the lighter $h$ boson but also the heavier
$H$ and $A$ particles are produced  in the little cascades, leading to very
clear peaks for the light and the  heavy neutral Higgs bosons. Again, in the
background spectrum, only the $Z$ boson peak can be observed. Both this scenario
and the previous one feature squarks lighter than the gluino [which has a mass  
around 1 TeV];  from the figures, it is clear  that this condition strongly 
enhances the visibility of the neutral Higgs boson signals. \s

Conservatively estimating the significance of the peaks as  ${\cal{S}}=S/ \sqrt{B}$,
where $S$ is the number of events in the peak and $B$  the number of events
``below" the peak  [i.e. below a hand--drawn line between the two shoulders of
the peak], we obtain a 5$\sigma$ significance for the $A,H$ peak in Sc4. With
30 fb$^{-1}$ of integrated luminosity, enough data are collected to  observe a
5$\sigma$ excess also for Sc3.  Sc1 and Sc2 will require a more detailed study
and/or a larger integrated luminosity.\s

This simple analysis of the neutral Higgs bosons produced in the SUSY cascades
shows that this alternative production mechanism looks very promising. The 
Standard Model backgrounds can be efficiently suppressed by the selection
criteria outlined above. By studying the $b\bar{b}$ invariant mass  spectrum,
we have shown that for all the considered scenarios  the neutral Higgs bosons
are visible after a few years of low luminosity running of the LHC collider. \s

This has been exemplified in the case where $M_A=150$ GeV and $\tb=5$. However,
as discussed in section 2, the dependence of the signal cross sections times
branching ratios on the parameter $\tb$ is rather weak in the cascade decays
and we can therefore extrapolate this conclusion to all values of this
parameter. For the Higgs mass range that can be probed in these processes, 
we are limited by two factors: \s

$(i)$ In the low mass range, $M_A \gsim 90$  GeV, we get rather close to the
$Z$ boson peak and the signal and  background overlap; however, the cross
section times branching ratios for the signal can be rather large in this 
phase--space--favoured case and the signal peak can be much larger than the
background peak. Moreover, by normalizing the $Z \rightarrow b\bar{b}$ 
distribution to the $Z \rightarrow ee, \mu\mu$ distribution, 
any excess can be clearly established. \s

$(ii)$ For larger $M_A$ values, the $A,H$ bosons become too heavy and 
phase space becomes penalizing [the branching ratios for the decays of heavier
chargino/neutralino states into the lighter ones and the $H,A$ bosons become
too small]. However, in many scenarios one can still have reasonable Higgs
production rates up to pseudoscalar masses of the order of $M_A=250$ GeV. 
Certainly, in Sc1--Sc4 discussed here, one can probe masses up to 
$M_A=200$ GeV, since there still are reasonable event rates, as shown in 
Figs.~1--4. \s

Therefore, it appears that the heavier MSSM Higgs bosons originating from
cascade decays  of squarks and gluinos can be detected at the LHC for any value
of $\tb$ and for pseudoscalar Higgs boson masses approximately in the mass
range between  100 and  200 GeV [the lightest $h$ boson can be detected in its
entire mass range] in a few years of LHC running with a moderate luminosity.

\newpage

\begin{figure}[htbp] 
\begin{center}
\vspace*{-.5cm}
\epsfig{file=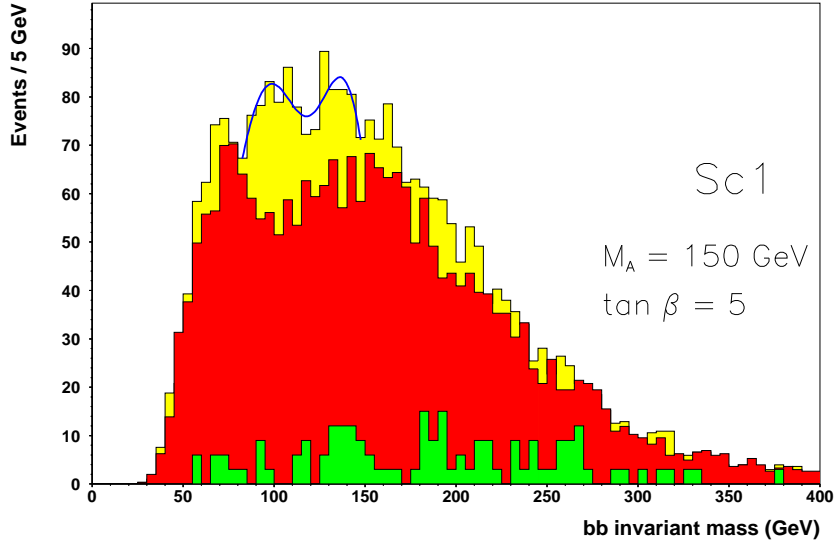,height=80mm}
\end{center}
\vspace*{-5mm}
\caption{\it Distribution of the $b\bar{b}$ invariant mass for the SUSY signal
events on top of the SUSY cascade background assuming scenario Sc1 (with 30
fb$^{-1}$ luminosity). At the bottom, the distribution for the Standard Model 
$t\bar{t}$ background is plotted. Two rather small peaks can be seen in the signal,
corresponding to  the presence of $h$ and $A,H$ bosons originating from the big
SUSY cascades.}
\label{fig:final1}
\vspace*{-5mm}
\end{figure}
\begin{figure}[htbp]
\begin{center}
\epsfig{file=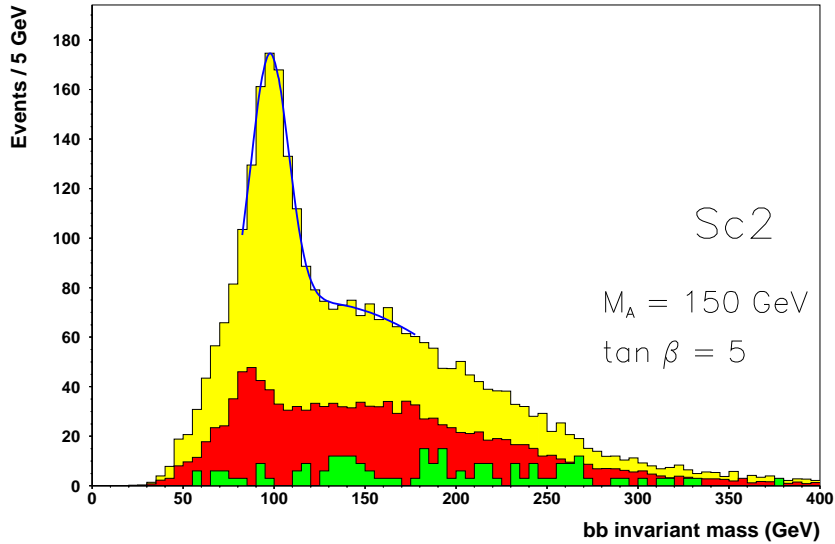,height=80mm}
\end{center}
\vspace*{-5mm}
\caption{\it Distribution of the $b\bar{b}$ invariant mass for the SUSY signal 
events on  top of the SUSY cascade background for scenario Sc2 (30 fb$^{-1}$).
The SM $t\bar{t}$ background is also shown. A large signal peak is visible, corresponding to the $h$ bosons that are
abundantly produced in the little cascades; also a much smaller and broader 
peak can be seen, signaling the presence of $A$ and $H$ bosons coming from  the
big  cascades.} \label{fig:final2}
\vspace*{-55mm}
\end{figure}

\newpage

\begin{figure}[htbp] 
\vspace*{-.5cm}
\begin{center}
\epsfig{file=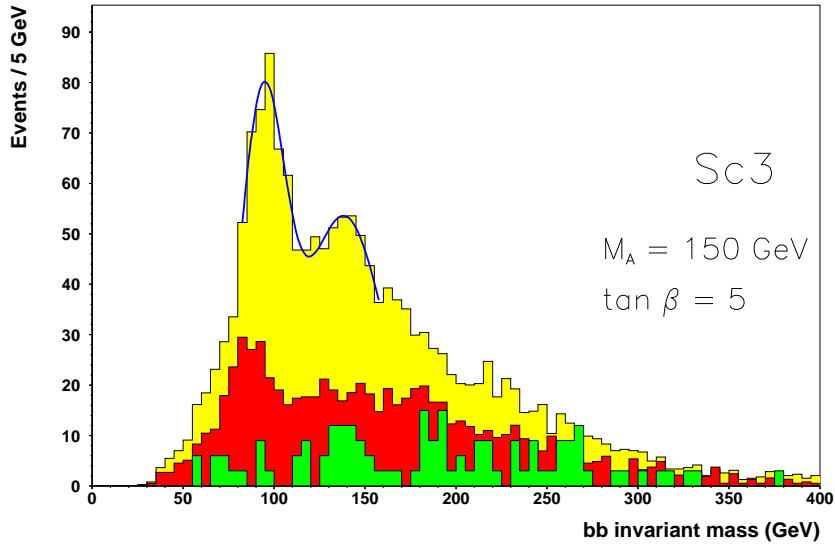,height=80mm}
\end{center}
\vspace*{-5mm}
\caption{\it Distribution of the $b\bar{b}$ invariant mass for the SUSY signal
events on top of the SUSY cascade background for scenario Sc3 (30 fb$^{-1}$). 
The SM $t\bar{t}$ background is also shown. Two peaks can be distinguished, a 
large one corresponding to the $h$ boson  and a smaller one corresponding to 
the $A,H$ bosons, all originating from the big cascades.}
\label{fig:final3}
\vspace*{-5mm}
\end{figure}
\begin{figure}[htbp] 
%\vspace*{-1cm}
\begin{center}
\epsfig{file=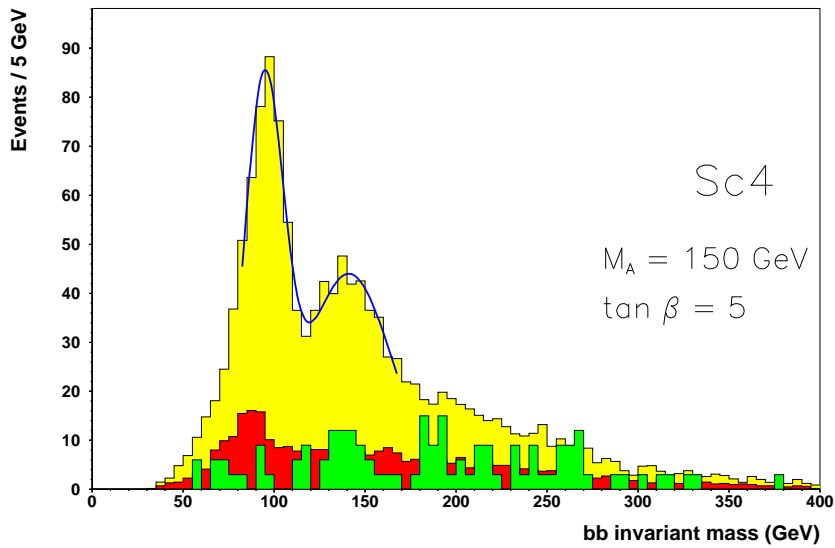,height=80mm}
\end{center}
\vspace*{-5mm}
\caption{\it Distribution of the $b\bar{b}$ invariant mass for the SUSY signal 
events on top of the SUSY cascade  background for scenario Sc4 (30 fb$^{-1}
$). The SM $t\bar{t}$ background is also shown. Two 
peaks are clearly visible, corresponding to the $h$ and $A,H$ Higgs bosons, 
all being produced in the little cascades.}
\label{fig:final4}
\vspace*{-55mm}
\end{figure}
\newpage

\subsubsection*{3.2 Charged Higgs bosons}

After having demonstrated the potential for observing the neutral Higgs bosons
in the  SUSY cascade decays, we will now study the charged Higgs production
through this mechanism. Since in all considered scenarios, $M_A \simeq 150$ GeV,
the charged Higgs boson mass is close to the top quark mass, $M_{H^\pm} \simeq
170$ GeV, and top quark decays into bottom quarks and $H^\pm$ bosons will be
strongly suppressed by phase space. For the same reason, the charged Higgs
decay mode to $t\bar{b}$ will be suppressed and the $H^{\pm}$ particles will
decay into $\tau \nu$ final states with a branching ratio of 95\% [some small
fraction would also decay into $cs$ final states for low values of $\tb$]. 
This enforces our choice of the  decay mode of investigation to be
$H^{\pm}\rightarrow \tau^{\pm} \nu_{\tau}$. \s

We will select events containing a hard $\tau$--jet plus additional hard jets
(often $b$-jets) accompanied by a large amount of missing energy due to the
presence of the lightest neutralinos and the neutrinos. Since there are numerous
sources of missing energy in the cascade decays, an invariant  mass
reconstruction of the charged Higgs using the $\tau \nu_{\tau}$ mode is not
possible. Also, in the $t\bar{b}$ decay mode, if the branching ratio had been
sizeable,  a reconstruction of the 4--jet system [2 $b$--jets + 2 jets from the
$W$ boson]  would have suffered from a huge combinatorial background in the
environment of  the cascade decays.\s

Though deprived from the powerful tool of invariant mass reconstruction, the 
hadronic $H^{\pm}\rightarrow \tau^{\pm} \nu_{\tau}$ mode has a special feature
that can be exploited in distinguishing  it from the main background decay $W^{\pm}\rightarrow
\tau^{\pm} \nu_{\tau}$: the tau polarization effects for both channels
are quite different \cite{roy}. \s

Indeed, due to the scalar nature of the decaying $H^+$ particle, the $\tau^+$
lepton from its decay is produced in a left--handed polarization state. In the
simplest scenario of $\tau^+ \ra \pi^+ \bar{\nu}$ decay, the right--handed
$\bar{\nu}$ is preferentially emitted in the direction opposite to the $\tau^+$
in the tau rest frame, so as to preserve the polarization. On the contrary, in
the background, the $\tau^+$ from the $W^+ \ra \tau^+ \nu$ decay is produced in
a right--handed state due to the vector nature of the $W^+$ boson, forcing the
$\bar{\nu}$ from $\tau^+ \ra \pi^+ \bar{\nu}$ to be emitted in the same
direction as the $\tau^+$ in the tau rest frame.  Therefore, harder pions are
expected from the signal than from the $W$ bosons in the background.  By
selecting events in which the fraction of the $\tau$-jet transverse momentum
carried by the charged pion is large, the SM and SUSY background involving $W$
bosons can be suppressed  relative to the signal. The tau leptons in the
background coming from  either $\tilde{\tau}$ and $Z$ decays or the decays of
the MSSM neutral Higgs bosons [in particular $H$ and $A$]  can however not be
suppressed this way. \s

Thus, as a signature for the charged Higgs boson in the SUSY cascade decays,
we will look for highly energetic $\tau$--jets in which the leading
charged particle carries most of the momentum.\s

For the simulations, we have again used {\tt ISASUSY 7.58} \cite{ISASUSY} for
the calculation of the SUSY particle spectrum  and the {\tt HERWIG 6.4}
\cite{HERWIG} Monte--Carlo event generator for the signal and background
generation. This generator takes into account the spin correlations between the
production and decay of all heavy fermions,  which in our case is important for
the top quarks in the $t\bar{t}$ production and SUSY particles in the cascades.
To properly account for the tau polarization, the generator was interfaced with
the {\tt TAUOLA} \cite{tauola} package. \s

We simulate again the four SUSY scenarios Sc1--Sc4 described in section 2, with
production cross sections times branching ratios as given in Fig.~5. For the SM
background, again only the $t\bar{t}$ process was considered. Also the $W$ +
$n$--jet background will be important, but we have not included it since the
standard Monte--Carlo generators do not give a realistic jet transverse energy 
spectrum for these processes, especially in the high transverse energy region.
We are, however, convinced that any selection criterion that is able to 
suppress the $t\bar{t}$ background, will also successfully reject the $W$ +
$n$--jet background. Since the kinematical distributions of the SUSY cascades
and the $t\bar{t}$ events are the same as in the neutral Higgs boson case, we
can refer to  Figs.~\ref{fig:distr1}--\ref{fig:distr4} and the related
discussion on the selection strategy to suppress the SM background. We will
adopt here the same selection criteria. \s

After eliminating the SM backgrounds, there still remains the more difficult
task to discriminate the $H^{\pm}$ bosons from the other particles in
the SUSY cascade decays. Tau--leptons in the cascade backgrounds originate
mostly from $W$ and  $Z$ bosons. Taus coming from these particles will have a
somewhat softer spectrum than the ones coming from a 170 GeV charged Higgs
boson. Therefore we impose a lower limit on the transverse energy of the
$\tau$-jet around the $W$ mass, $E_T^{\rm \tau-jet}$  $>$ 80 GeV. \s

As a measure of the hardness of the charged hadron in the $\tau$--jet,  we show
in Fig.~\ref{fig:charfinal} the distribution of the fraction of the transverse
energy $E_T^{\rm \tau-jet}$ that is carried by the one-prong charged particle
in the jet. The full line corresponds to the SUSY signal [i.e. the cascade
decays including the charged Higgs bosons] while the dashed line corresponds 
to the SUSY backgrounds [i.e. the same cascades where the decays of the heavier
charginos/neutralinos to charged Higgs bosons and lighter charginos/neutralinos
are forbidden]. Because of the polarization effects as explained above,  this
distribution peaks at high values for tau leptons coming from scalar charged
Higgs bosons, while peaking at low values for tau leptons coming from
vector--like $W$ bosons.  Requiring that at least 75\% of the transverse energy
of the $\tau$--jet be carried by the charged hadron, we can enhance  the
visibility of the signal with respect to the SUSY background.\bigskip

\begin{figure}
\begin{center}
\epsfig{file=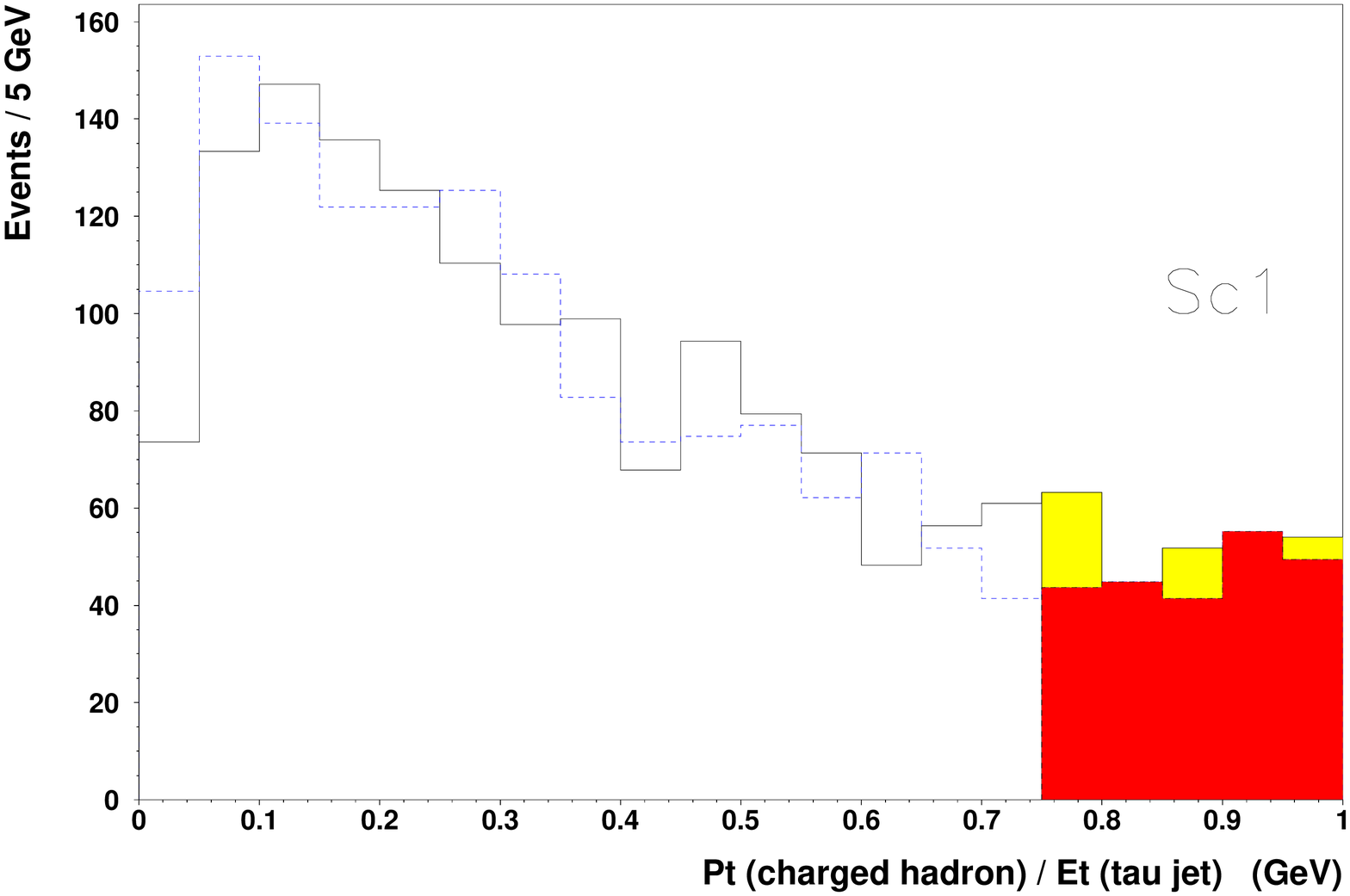,height=85mm,width=75mm}
\epsfig{file=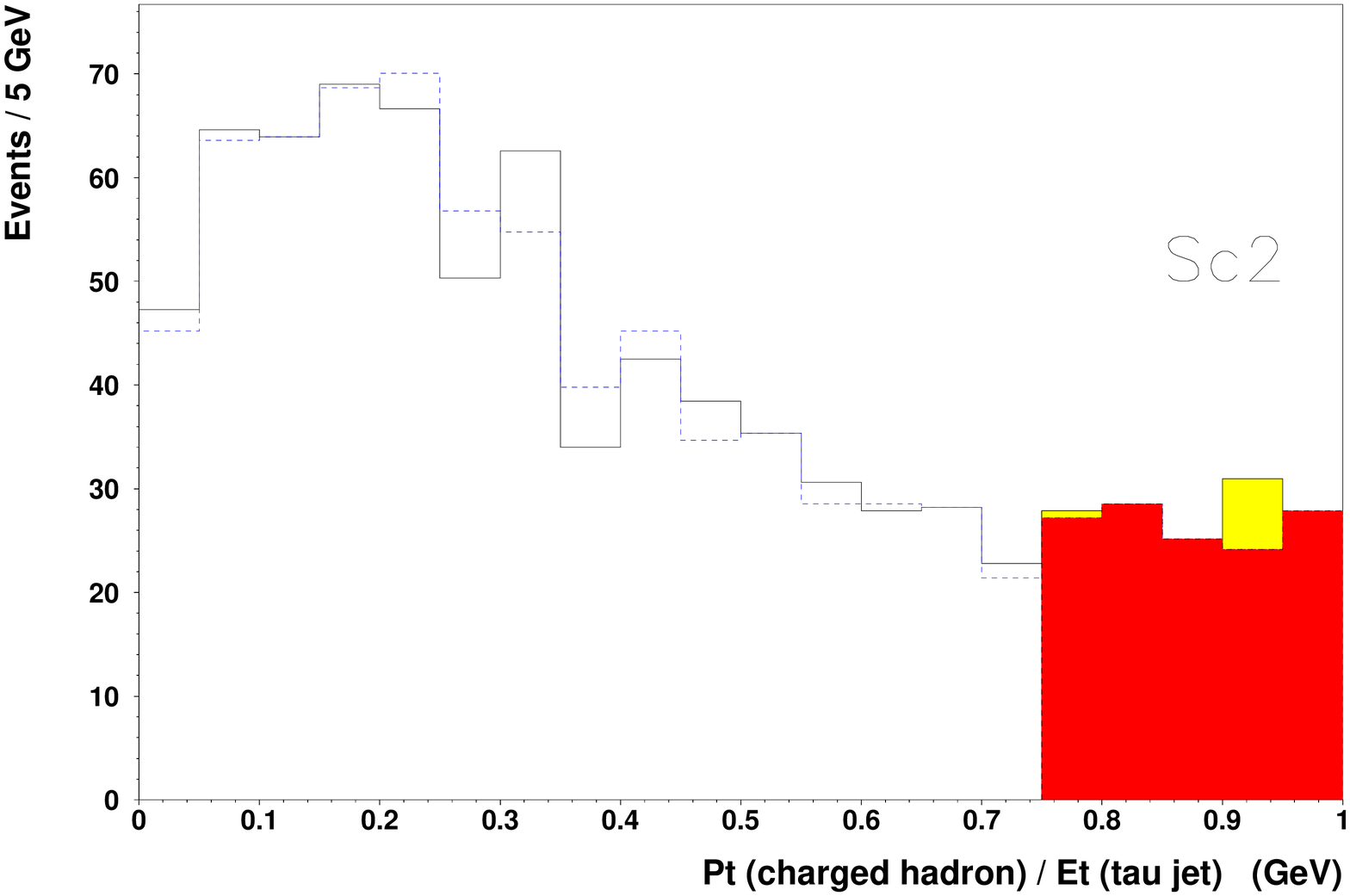,height=85mm,width=75mm}
\end{center}
\begin{center}
\epsfig{file=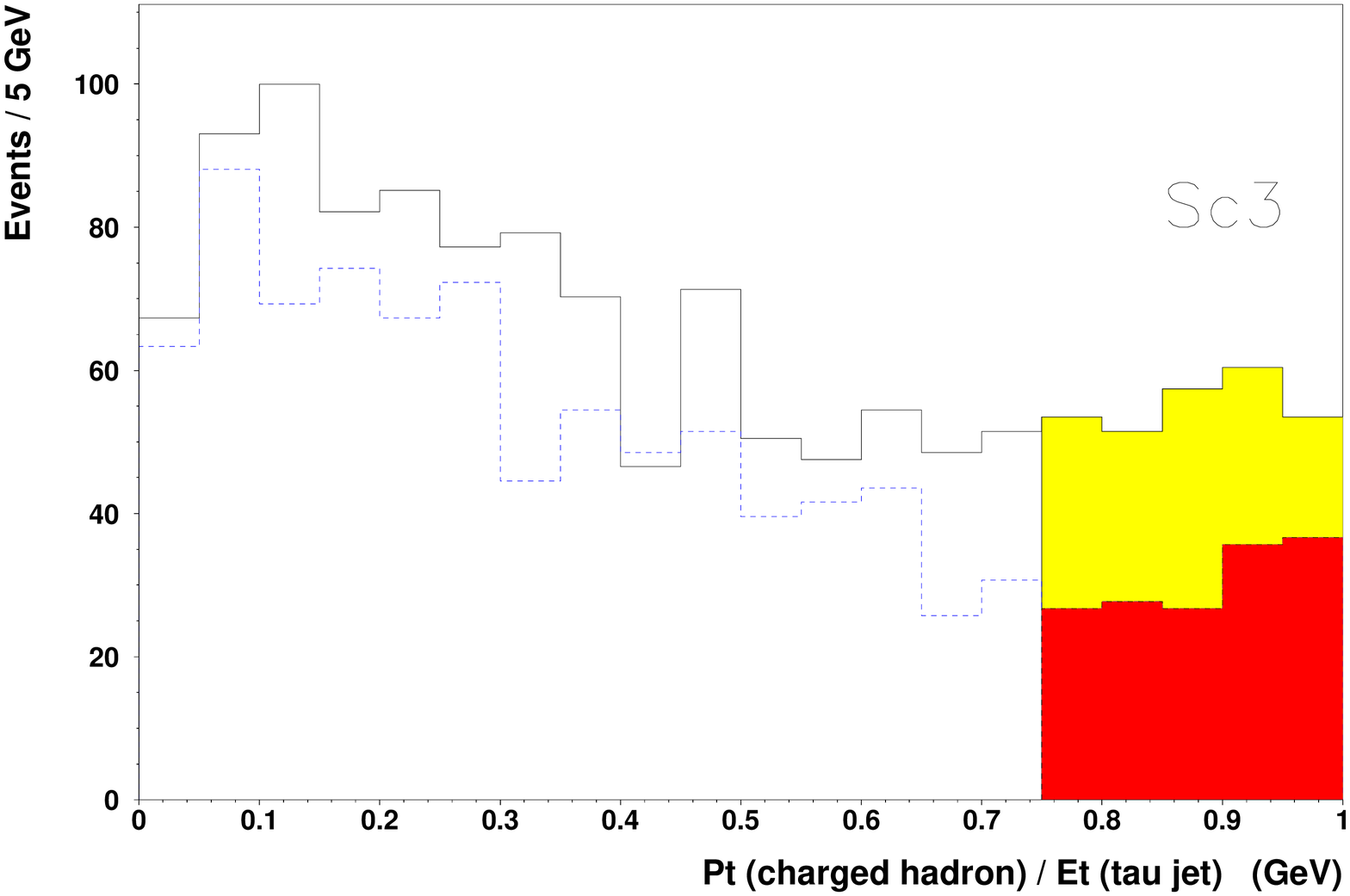,height=85mm,width=75mm}
\epsfig{file=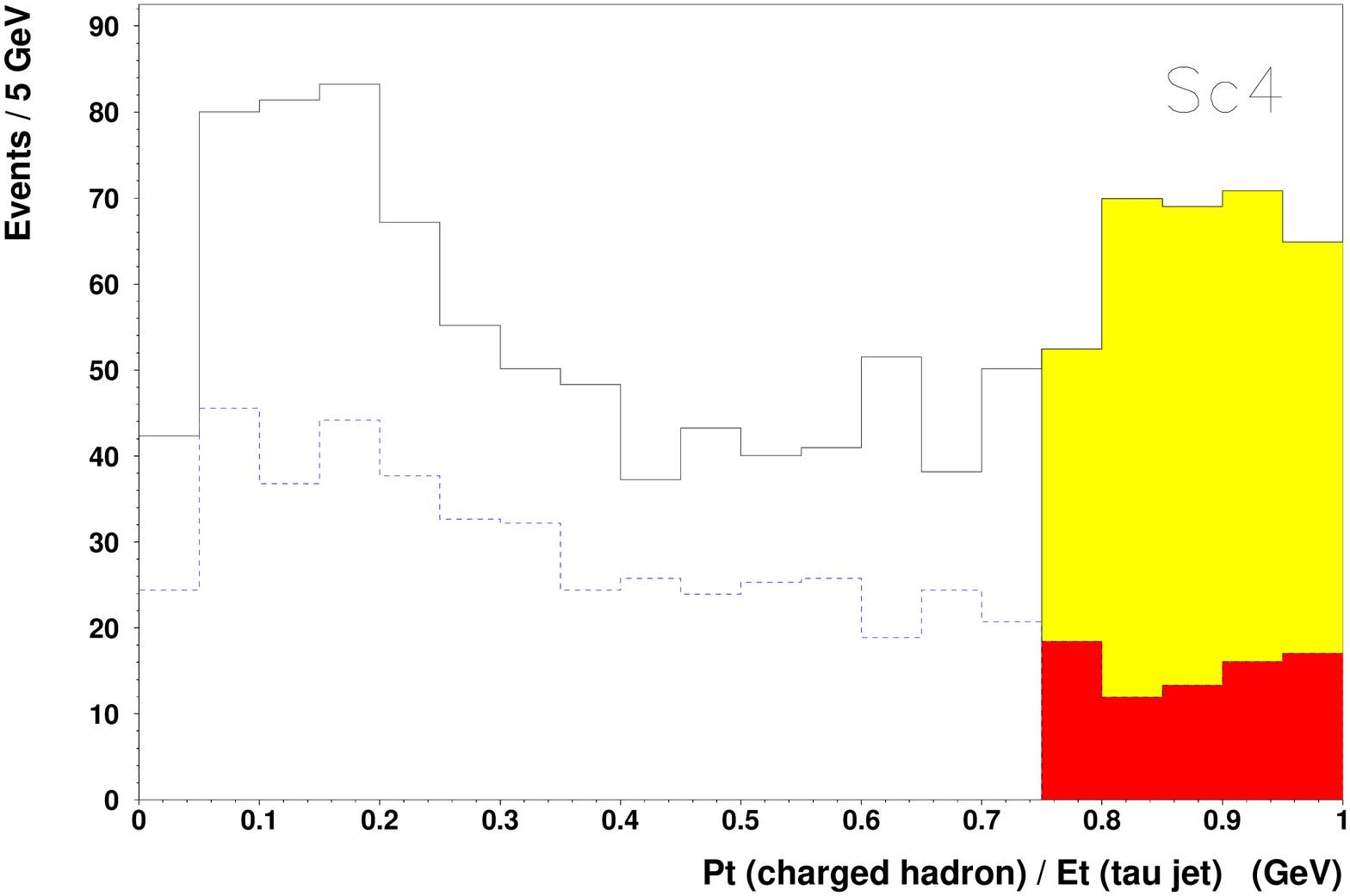,height=85mm,width=75mm}
\end{center}
\caption{\it Distribution of the fraction of the $E_T^{\rm \tau-jet}$ that is
carried by the  charged particle in the jet. The full line represents the
signal  while the dashed line represents the SUSY background. The shaded areas
contain the events left after applying the selection cuts, for  an integrated
luminosity of 30 fb$^{-1}$. The charged Higgs signal is clearly visible in
 Sc3 and Sc4, but not visible in  Sc1 and Sc2.}
\label{fig:charfinal}
\end{figure}

These considerations lead us to the following selection criteria in order to
distinguish between signal and background events: 

\begin{itemize}

\item[$i)$] The events should contain at least five jets.  

\item[$ii)$] The hardest jet in the event should have $E_T$ $>$ 300 GeV.  

\item[$iii)$] The transverse missing energy $E_T^{\rm miss}$ in the event 
should be larger than 200 GeV.  

\item[$iv)$] We require exactly one hadronically decaying tau (1-prong),  i.e.
we demand a narrow jet within $|\eta| \, < \, 2.5$,  which should contain  a
hard charged track of $p_T \, > \, 5$ GeV in a cone of $\Delta R$ = 0.15 rad
around the calorimeter jet axis, and it should be isolated, meaning no charged
tracks with $p_T \, > \, 2$ GeV are allowed in a cone of $\Delta R$ = 0.4 rad
around the axis.  

\item[$v)$] The $E_T$ of the $\tau$--jet, defined as the $E_T$ reconstructed in
a cone of $\Delta R$ = 0.4 rad around the jet axis, should be above 80 GeV.

\item[$vi)$]  More than 75\% of the $\tau$--jet transverse energy should be
carried by the charged track, i.e. we impose the cut  $p_T^{\pi}/E_T^{\rm 
\tau-jet}$ $>$ 0.75.  
\end{itemize}

Events that satisfy conditions ($i$)--($iii$) can be efficiently triggered
on using the jet and missing--energy triggers.  \bigskip

In Fig.~\ref{fig:charfinal}, the shaded areas represent the number of events
left after all selection cuts, for the signal and the SUSY background. The
charged Higgs boson signal is clearly visible in  Sc3 and Sc4, but not in  Sc1
and Sc2. This is because the production cross section for these scenarios are
about 10 times smaller than for  Sc3 and Sc4, as can be seen from Fig.~5. \s

Clearly, the evidence for the charged Higgs boson is not as convincing as the
invariant mass peaks we obtained in the neutral Higgs boson case. However, in
the MSSM, the neutral and charged Higgs boson masses are not independent: since
they are are connected through the relation $M^2_{H^\pm}=M^2_A + M^2_W$ [which
is valid at leading order but is not strongly affected by radiative
corrections], the observation of the neutral Higgs bosons can give definite
information  on the charged Higgs boson mass. If one then observes an excess in
$\tau$--jet events after a selection as illustrated above, one can be rather 
confident that this excess originates from the production of charged Higgs
particles\footnote{Note that for relatively light charged Higgs bosons which 
can be produced through top quark decays, the situation can be slightly more
complicated. Although the production rates of $H^\pm$ bosons originating from
cascade decays of squarks and gluinos  can be enhanced as discussed in section
2.4, one has to deal with the additional events coming from direct production
of $t\bar{t}$ pairs. More detailed analyses of this topic will be needed.}.

%{\bf Some numbers to be filled in } ...  meaning that a 5$\sigma$-observation
%of $H^{\pm}$ bosons produced in SUSY cascade decays could be made with about 2
%years of high luminosity data of the LHC, assuming the above physics scenario
%and provided the SUSY background processes are well understood.  

\newpage

\subsection*{4. Conclusion} 

We have analysed the detection prospects of the neutral and charged Higgs
particles of  the MSSM in single Higgs final states obtained through the
cascade decays of the heavy squarks and gluinos that will be copiously produced
at the LHC. We have discussed in detail the case of the production through the
cascades involving the heavier charginos and neutralinos [originating from the
decays of the strongly interacting SUSY particles] into  the lighter charginos
and neutralinos and a Higgs boson $\Phi$. This can occur  through the ``big
cascades", $\chi_{3,4}^0, \chi_2^\pm \to \chi_{1,2}^0, \chi_1^\pm + \Phi$, as
well as through the ``small cascades", $\chi_{2}^0, \chi_1^\pm \to
\chi_{1}^0+\Phi$, or through both cascades, depending on the considered
scenario. For illustration, we have selected four scenarios which are expected
to cover most of the situations which can occur in the MSSM: squarks either
lighter or heavier than gluinos, and light charginos and neutralinos, either
of the gaugino-- or higgsino--type. \s

We have shown that in these scenarios and with not too heavy Higgs
particles, with masses $M_\Phi \lsim 200$--250 GeV, the cross sections times
branching ratios for the production of at least one neutral or charged MSSM
Higgs boson can be substantial at the LHC\,\footnote{At the Tevatron, because of
the reduced kinematical reach for squarks and gluinos, one would have access to
these final states in much more limited  areas of the parameter space. However,
the production of the neutral Higgs boson through the little cascade, $\chi_2^0
\to \chi_1^0 +h$, might be kinematically possible and could be searched for
with a high integrated luminosity.}, resulting in large numbers of signal
events with an integrated luminosity of $\int {\cal L} \gsim  30$ fb$^{-1}$,
which is expected to be collected after a few years of running.  \s

We have performed a full Monte--Carlo simulation  that takes into account the
various signals as well as the SM and SUSY backgrounds and includes a fast
simulation of some important aspects of the responce of the CMS detector at the
LHC [similar conclusions are expected to be drawn for the ATLAS detector]. We 
have shown that these final states can indeed be detected in some 
representative MSSM scenarios. In particular, the heavier CP--even $H$ and the
pseudoscalar $A$ bosons can be observed in the region of parameter space $M_A
\sim 150$ GeV and $\tb\sim 5$, which cannot be probed  by searches using the
standard Higgs production channels. Therefore, the search for Higgs bosons 
via the cascade decays could be complementary to the standard searches. \s

\begin{figure}
\begin{center}
\hspace*{-.7cm}
\epsfig{file=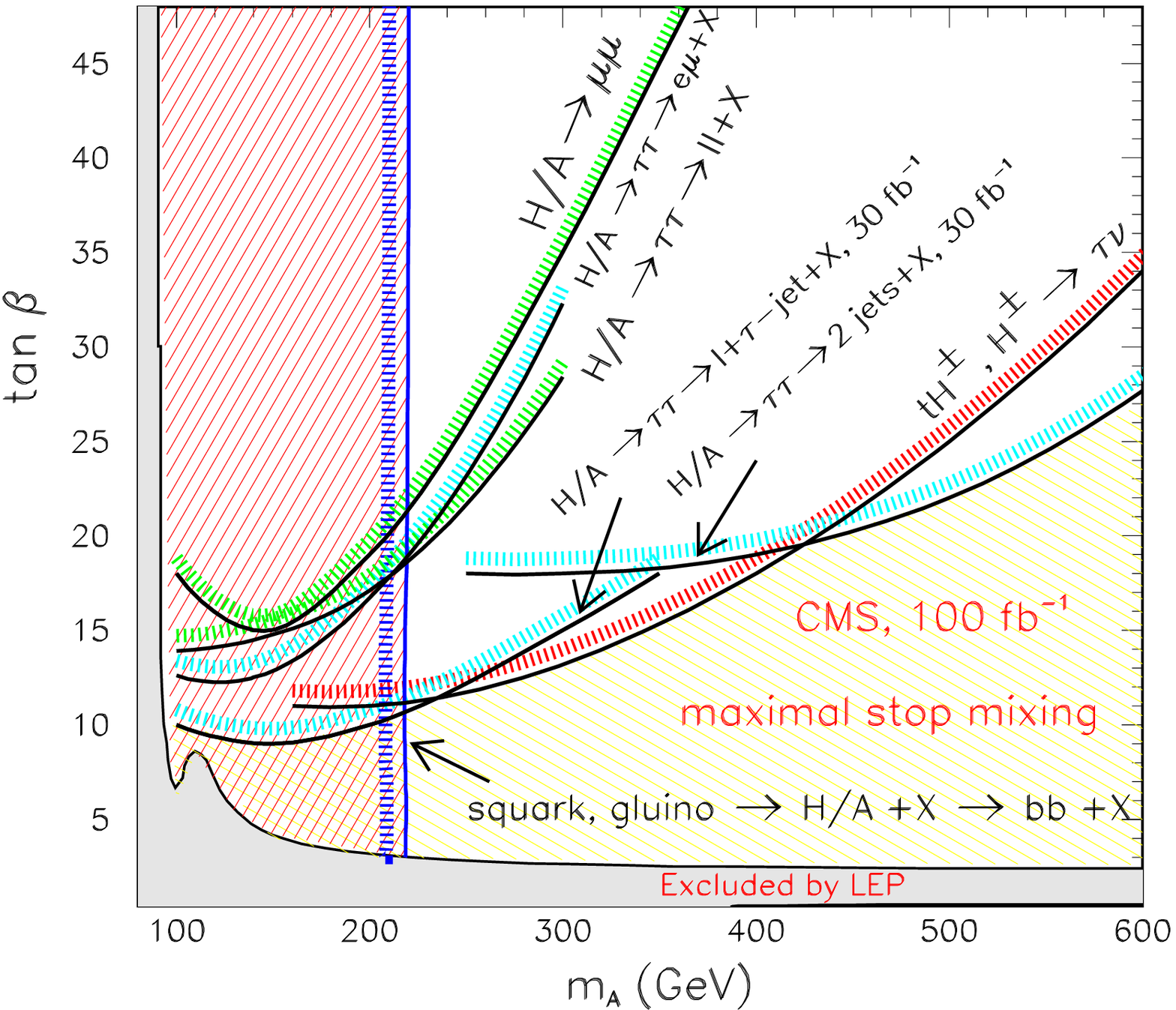}
\end{center}
\caption{\it The areas in the $(M_A, \tb)$ parameter space where the MSSM Higgs
bosons can be discovered at CMS with an integrated luminosity of 100 fb$^{-1}$.
Various detection channels are shown in the case of the standard searches
for the maximal stop mixing scenario. The right--hatched and cross--hatched 
regions show the areas where only the lightest $h$ boson can be observed in 
these production channels. The left--hatched area is the region where the 
heavier CP--even $H$ and pseudoscalar $A$ bosons can be observed through the 
(big) cascade decays of squarks and gluinos in  Sc3 with $M_2=350$ GeV.
A similar contour plot can be obtained in Sc4, where only the
little cascade is at work.} 
\label{fig:final}
\end{figure}

This is exemplified in Fig.~17, in which we show the areas of the $(\tb , M_A$)
parameter space where the MSSM Higgs particles can be detected by the CMS
collaboration with an integrated luminosity of $\int {\cal L} \simeq 100$
fb$^{-1}$. Besides the usual areas where the Higgs bosons can be probed in
SM--like production and decay processes, we exhibit the range of $M_A$ values
where one can  detect, in addition, the $A,H$ and $H^\pm$ bosons via the 
cascade decays in a  specific MSSM scenario [namely Sc3, with the 
value of the parameter $M_2$ fixed to $350$ GeV]. The  heavier neutral $A,H$ 
particles can be probed for masses up to $M_A \sim 220$ GeV for the 
entire range of $\tb$ values [as we have shown in our analysis, the cross 
sections times branching ratios depend only mildly on the value of this 
parameter], while the charged Higgs particle can be observed for masses up to
$M_H^{\pm} \sim 200$ GeV [the lighter $h$ boson can of course be detected in  the
entire ($\tb, M_A)$ plane in this scenario]. As can be seen, the additional
channels can fill some holes of the  parameter space where these heavier Higgs
particles cannot be observed  through the standard  processes. \s

We have also discussed, though only at the level of  production cross sections
times decay branching fractions, MSSM Higgs boson production through the
decays of heavier third--generation squark into the lighter ones, as well as the
production of $H^\pm$ bosons from the decays of top quarks originating from
$\tilde{g}, \tilde{q}$ cascade decays. While the rates for Higgs boson 
production from direct decays of third--generation squarks are in general 
smaller than in the previous case, they can be substantial for large $\tb$ 
values in the case where top quarks originating from SUSY cascades decay  into
$H^\pm$ bosons. These additional channels can therefore increase the  yield for
Higgs particles at the LHC.  We did not  attempt to perform a Monte--Carlo
simulation to verify whether the final states in these channels can indeed
be detected in a realistic situation. \s

The present study does not pretend to be exhaustive. To cover the entire
possibilities for MSSM Higgs boson production  through SUSY particle cascade
decays, many more studies, in particular detailed experimental simulations,
are required  in order to fully cover this rather complicated
subject. Here, we  simply made a preliminary investigation in some selected 
scenarios, which indicates that the detection of Higgs particles in these
processes is not at all hopeless and might even complement the searches 
through the intensively studied SM--like processes in some favourable cases. \s

We stress again that these cascade processes are important not only because
they represent a new source of Higgs bosons, but also because they will be very
useful to measure the couplings of supersymmetric particles to Higgs bosons,
which would be an essential ingredient to reconstruct the SUSY Lagrangian at
low energies and to probe the theory at very high scales.\s 

In addition,  if the branching ratios are sufficiently large, the decays of heavy
neutralinos  into $b\bar{b}$ pairs through Higgs boson intermediate states, 
$\chi_{3,4}^0 \rightarrow h,H,A +\chi_1^0 \rightarrow b\bar{b} + \chi_1^0$, 
contain information to reconstruct the masses of the heavier neutralinos and
could as such help significantly in identifying the supersymmetric  particle
spectrum at the LHC.\s

Hence, on their own merit, SUSY particle cascade processes  involving Higgs 
bosons deserve detailed and dedicated studies in the future. \newpage

\nn {\bf Acknowledgements:} \s

\nn We thank Daniel Denegri, Yann Mambrini and Luc Pape for  discussions. 
A.~Datta would like to acknowledge the French MNERT fellowship, the CNRS for
partial support and the LPMT for the hospitality accorded to him when this work
was initiated; he is supported by a DOE grant DE-FG02--97ER41029. M.~Guchait
thanks the CERN Theory Division, where part of this work was  performed.
F.~Moortgat is research assistant of the Fund for Scientific Research
(FWO-Vlaanderen), Belgium.

%\newpage

\end{document}